\documentclass[12pt]{article}
\pdfoutput=1
\usepackage{jheppub}
\usepackage{graphicx}
\usepackage{epsfig}
\usepackage{epstopdf}
\usepackage{amsfonts}
\usepackage{amssymb}
\usepackage{amsmath}
\usepackage{bm}% bold math
\usepackage{hyperref}
\usepackage{mathrsfs}
\usepackage{bbm}
\usepackage{cancel}
\usepackage{multirow}
\usepackage{subfigure}
\usepackage{color}
\usepackage{breqn}
\usepackage[numbers,sort&compress]{natbib}
\def\be{\begin{equation}}
\def\ee{\end{equation}}
\def\bea{\begin{eqnarray}}
\def\eea{\end{eqnarray}}

\title{Primordial Gravitational  Waves Spectrum in the 
Coupled-Scalar-Tachyon Bounce Universe}
\author[a]{Nan~Zhang,}
\author[a]{Yeuk-Kwan~E.~Cheung}

\affiliation[a]{Department of Physics,
Nanjing University,\\
22 Hankou Road, Nanjing, China 210093}
\emailAdd{zn95330@outlook.com}
\emailAdd{cheung@nju.edu.cn}

\abstract{
We derive in detail the equations of motion for the tensorial  modes of primordial 
metric perturbations in the  Coupled-Scalar-Tachyon Bounce Universe.
We  solve for the gravitational wave equations  in  the pre-bounce contraction
and the post-bounce expansion epochs.
To   match  the solutions of the  tensor perturbations, we idealise the bounce process 
yet retaining the essential physical properties of the bounce universe. 
We  put forward two matching conditions: one ensures the continuity of the gravitational
wave functions and the other respects the symmetric nature of the bounce dynamics. 
The matching conditions  connect  the two independent modes of gravitational waves 
solutions before and after the bounce.  We further analyse the scale dependence 
and time dependence of the gravitational waves spectra in the bounce universe and 
compare them with  the primordial spectrum in the single field  inflation scenario. 
We discuss the  implications to early universe physics and present model independent 
observational  signatures extracted from  the bounce universe. 
}

\keywords{Bounce Universe, Primordial Density Perturbations, Metric Tensor Perturbations, Gravitational Waves}

\begin{document}
\maketitle
\section{Introduction}
\label{sec:intro}

Encouraged by the direct detections  of the  gravitational waves in the spiralling blackhole and neutron star systems~\cite{Abbott:2016blz, Abbott:2016nmj, Abbott:2017vtc, Abbott:2017gyy, Abbott:2017oio, TheLIGOScientific:2017qsa, GBM:2017lvd}
exactly one century after the prediction of their existence by Einstein,  
a number of observational effects  are
being set up to detect  primordial gravitational waves  originated from the ``birth''
of our currently observed  universe,   hence called the  ``primordial
gravitational waves'' or ``relic gravitational waves''%
~\cite{AmaroSeoane:2012je, AmaroSeoane:2012km, Ade:2014xna, Ade:2014gua, Luo:2015ght, Gong:2014mca, Cai:2016hqj, Li:2017drr, Ade:2018gkx}.
This shall  open up a new window in the study of early universe cosmology and high
energy physics,  which, in turn,  have profound impacts on our understanding of the physics  of the  earliest epoch of our  universe:
Gravitational waves decouple  from the hot plasma
at thermal equilibrium  at  $T_{decouple}\sim T_{Planck}$,
much earlier than all other particles and propagate freely afterwards.
Gravitational waves thus capture the snapshots of the stochastic background of the
 universe which encode invaluable quantum-gravity information of this  earliest epoch
  of  our universe. They encode information of physical processes at extremely high
  energy  otherwise unattainable  by particle accelerators on Earth and will 
  perhaps  shed light on  the origin of matter.
In short gravitational waves physics from the earliest epoch of the universe  thus   nicely compliment  and greatly enhance the existent  discriminating  power of
cosmology data brought about by the current era of precision cosmology.

Bounce universe -- in which the early universe is proposed to have a phase of 
matter-dominated contraction prior to the usual ``Big Bang'' and  the ensuing  
(but not necessarily exponential) expansion -- has emerged over the past decade or 
so to be a viable alternative to inflation.
The discovery that the  primordial spectrum of density perturbations generated in the contraction phase being (albeit naively)
scale invariant in a seminal paper by D.Wands~\cite{Wands:1998yp} 
(See also~\cite{Finelli:2001sr, Gratton:2003pe}.) 
has since then encouraged a continual stream of research effort. 
Reviews have been  written over the years
to timely document  this line of research activities~\cite{Novello:2008ra, 
Liu:2010fm, Brandenberger:2012zb, Battefeld:2014uga, Brandenberger:2016vhg, 
Yeuk-KwanEdnaCheung:2016zra,  Nojiri:2017ncd}. 

In this project we are going to obtain a complete solution to the gravitational 
waves equations  in  a simplified bounce universe model, taking careful account 
of two possible matching conditions~\cite{Deruelle:1995kd, Durrer:2002prd}.  
We also  compute  the gravitational waves spectrum generated from  this  bounce universe and compare it with the gravitational waves spectrum in a single 
field inflation model, recently obtained in~\cite{Alba:2015cms}. 

The bounce universe model, the gravitational waves spectrum of which we will 
analyse in detail,   was  derived from low energy effective action of string 
theory.  It  built on the idea of non-BPS D-branes and anti-D-branes 
annihilation~\cite{Dvali:1998pa, Dvali:2001fw, Burgess:2001fx}. 
The tachyon effective potential had been  used, independently, by 
Gibbons~\cite{Gibbons:2002md} and  Sen~\cite{Sen:2003mv} to model inflation. 
 
By introducing a low energy coupling between the tachyon and Higgs fields living on 
the  D-branes we were  able to make a model of a  bounce universe, 
called the CST bounce universe. Not only can  the CSTB model   
generate enough inflation to solve the ``flatness,'' ``homogeneity,''  and ``horizon'' 
problems of the Big Bang model~\cite{Li:2011nj, Cheung:2016oab}, it can also address the ``cosmic singularity''  problem  because  the universe  could be  dynamically 
stabilised at a non-zero  minimal scale. 
In~\cite{Li:2013bha} we obtained  the 
primordial spectrum  of the scalar  perturbations and established   its scale 
invariance  and  its  stability under  time evolution.   
In the latter series of 
work~\cite{Li:2014era,Cheung:2014nxi,Cheung:2014pea, 
Yeuk-KwanEdnaCheung:2016zra,Vergados:2016niz}  we undertook 
a model independent study of dark matter creation and evolution in a bounce universe. 
We  were able to extract a characteristic curve relating dark matter coupling and 
dark matter mass should the production  of the dark matter take place  via an 
out-of-thermal-equilibrium route in a bounce universe.

This  paper is organised as follows.  In Section~\ref{sec:tensor-modes} we discuss the 
dynamics of a bounce universe model  built out of the tachyon and Higgs fields, 
 and their low energy effective coupling as much as it is needed to study the 
 dynamics of the tensorial modes of metric perturbations in the bounce universe. 
Even though the equation of curvature perturbations is simple and elegant in the 
 CSTB universe,  the equations  of tensor  perturbations 
   become more complex due to  a frictional  term. 
   In Section~\ref{sec:tensor-modes} 
we derive the equation of tensor modes by varying the Lagrangian of 
coupled scale and tachyon fields in  a perturbed  background metric.
In Section~\ref{sec:simplified_bounce}  
we  analyse  the behaviour  of  the tensor modes  as they  cross  the effective horizon. 
 After a qualitative discussion, we solve the tensor perturbations equations  
  in the  contracting and the  expanding  phases of CSTB universe
   and put forward two possible  matching conditions.
The   different power spectra arising  from the  continuous matching and the symmetric matching 
condition  are then presented.  
In Section~\ref{sec:power_spectrum} various observational signatures are computed   for the 
tensor modes from the bounce universe  and  their   implications   discussed. 
A side-by-side comparison with  the single-field  inflationary scenario is presented in Section~\ref{sec:comparison}; and  afterwards we conclude with a summary of results
and an outlook.  Pertinent resulted of the  GW spectrum  from single-field inflation recently 
obtained by Alba and Maldacena~\cite{Alba:2015cms} is 
reproduced  in Appendix A for easy reference and comparison. 
 %%%% 

\section{The dynamics of tensor perturbations in the CSTB universe}
\label{sec:tensor-modes}
A string cosmology model,  called the Coupled
Scale-Tachyon Bounce (CSTB) universe,  starts its life  from  the phase of
tachyon-matter-dominated contraction, onward to a  bounce at $H=0$  and
then into   a  phase of  post-bounce expansion from which our observed universe 
is produced.  A scale invariant and  stable spectrum of primordial matter density 
perturbations is produced during the contraction phase~\cite{Li:2013bha}. 
The spectrum of the scalar perturbations  is then evolved through the bounce; and 
its scale invariance and stability are  shown to be unaffected by the dynamics of 
the potentially strong gravity 
during the bounce process via an AdS/CFT analysis~\cite{Ming:2017dtm}. 
The rest of this paper is devoted to the computation of the  tensor modes of metric perturbations. The spectrum will  be evolved to the time of proton-electron
capture and obtain the  B-mode correlations in the CMB polarisations.
 The spectral index and its  running as well as the tensor-to-scalar 
ratio of the metric perturbations will also be presented.  

The primordial density perturbations generated during the
 contraction phase were  first found to be scale invariance by 
 Wands~\cite{Wands:1998yp} (See also~\cite{Finelli:2001sr, Gratton:2003pe}.).
 The horizon exit and reentry of these primordial density perturbations are indeed 
  similar to the analogous process in the inflationary scenario. 
  %especially when presented in the Hubble scale in conformal time. 
 However a closer look of the detailed dynamics reveals that each wavelength of the 
 perturbations in a bounce universe exits the  Hubble radius  at different times. 
 This makes each k-mode of the   perturbations pick up 
 an implicit time dependence which adds to the difficulty of model 
 building~\cite{Gratton:2003pe}. Care must be exercised when computing the 
 spectrum of primordial perturbations.  After taking careful  account of this implicit time 
 dependence the spectrum of CST bounce model is proven to be truly time independent and 
 scale independent~\cite{Li:2013bha}.

\subsection{Dynamics of the CST bounce universe}
In this section the  dynamics of the background  universe will be recalled as it 
is needed when we discuss the dynamics of the tensor modes inside the bounce universe. 
Just as the name coupled scalar-tachyon suggests, the CSTB model utilises  
a tachyon  field $T$ and its interaction  with a scalar (Higgs) field 
$\phi$~\footnote{%
In the string theory language, this class of scalar fields are open strings
stretching from one D-brane to anther and they play the roles of Higgs in the
low energy field theory.%
}.
The two-field potential creates a dynamical false vacuum when %the universe is
the tachyon $T$ is being  pulled up  its  potential hill  and locked at the  peak by the fast oscillations of the scaler field $\phi$ around the potential minimum of the latter. 
The minimal coupling of these two fields introduced in~\cite{Li:2011nj} makes
a bounce universe  possible.  CST bounce model utilises and extends the tachyon 
inflation models  first introduced  by Gibbons~\cite{Gibbons:2002md} and 
by Sen~\cite{Sen:2003mv}.

Analytically, the coupled scalar-tachyon bounce model can be described by an 
effective field theory,  the Langrangian density of which is comprised of
\begin{equation}
\label{L0}
\mathscr{L}_{CSTB}(T,\phi)=\mathscr{L}(T)+\mathscr{L}(\phi)-\lambda\phi^2T^2.
\end{equation}
In~(Eq.~\ref{L0}), $\mathscr{L}(T)$ stands for the  open string tachyon
 Lagrangian, $\mathscr{L}(\phi)$ is the Lagrangian of the
scalar  and $-\lambda\phi^2T^2$  a low-energy effective
coupling  between the tachyon  and the scalar,  
with  $\lambda$ denoting a dimensionless coupling constant. 
The tachyon  Lagrangian is written as (with the metric convention, $diag\{-,+,+,+\}$):
\begin{equation}
\label{lag tachyon} \mathscr{L}(T)= -V(T)\sqrt{1 + \partial_\mu
T\partial^\mu T},
\end{equation}
with $V(T)$ corresponding to the weakly attractive potential between a stack of D-branes and anti--D-branes~\cite{Dvali:1998pa, Dvali:2001fw,
Burgess:2001fx, Gibbons:2002md, Sen:2003mv}:
\begin{equation}
\label{V(T)} V(T)=\frac{V_0}{cosh(\frac{T}{\sqrt{2}})}.
\end{equation}
The tachyon Lagrangian describes the onset of an annihilation process of a
pair of D4-branes and anti-D4-branes in a closed universe~\footnote{
To avoid ghosts and violation of the null, weak, and strong energy conditions,  
we put our bounce universe model in a closed FRLW background and obey the soft bounce conditions~\cite{Borde:1993xh}.
}.
\begin{equation}
\label{lag phi} \mathscr{L}(\phi)=-\frac{1}{2}\partial_\mu
\phi\partial^\mu\phi-\frac{1}{2}m_{\phi}^2\phi^2.
\end{equation}
In the  string language, the scalar field $\phi$ can be simply
viewed as the distance between the two stacks of D-branes and
anti-D-branes.

The Friedmann equation is thus,
\begin{equation}
\label{eq:friedmann}
H^2=-\frac{1}{a^2}+\frac{8\pi G_{N}}{3}\rho~. 
\end{equation}
In the CST bounce universe  model, the energy density of the
isotropic and homogeneous background  universe is given by,
%where $\rho$ being  the total energy density  of  the universe at a given time:
\begin{equation}
\label{general energy density}
\rho=\frac{V(T)}{\sqrt{1-\dot{T}^2}}+(\frac{1}{2}m_{\phi}^2+\lambda T^2)\phi^2+\frac{1}{2}\dot{\phi}^2~,
\end{equation}
where all the fields  are  functions of time only. 
$8 \pi G_{N}\equiv {m_p}^{-2}$  will be used interchangeably in the rest of the paper. 
The  cosmic evolution of the universe  can be obtained, once the initial boundary 
conditions of the universe  are prescribed.
As we shall see the  prescription of boundary conditions in the bounce universe 
is  an integral part of the ensuing discussion on gravitational waves spectrum.

\textbf{Cosmic evolution before the bounce -- a matter dominated contraction phase: }
In a bounce universe model the universe is postulated to undergo a period of
matter-dominated contraction in which the universe comes into thermal contact to
restore causality, and hence addresses  the Horizon Problem.  
Casted in  a closed FRLW background, %~\footnote{
%To avoid violation of the null, 
%weak, and strong energy conditions~\cite{Borde:1993xh}.
%}
 the CST bounce universe undergoes  a contraction
phase dominated by the tachyon matter during   a reversal of tachyon
condensation~\cite{Sen:2004nf}.
As the universe contracts, each mode inside the universe gains energy. 
In particular the tachyon is being  pulled up its potential hill  
as a result of  its coupling with the Higgs field; and eventually, being dynamically  
locked at its false vacuum by the fast oscillations of the Higgs.  
The  stack of D-branes and anti-D-branes approach each other and 
this instability at the onset of co-annihilation is  reflected in the   
tachyon potential  as the tachyon  is approaching  its potential maximum. 
The  stack of D-branes and anti-D-branes  approach each other due to
the weakly attractive potential between the stacks of D-branes and
anti--D-branes~\cite{Dvali:1998pa, Gibbons:2002md, Sen:2003mv, HenryTye:2006uv}.
The energy of the universe during this process  is progressively 
dominated by this vacuum energy.
During the reversal of tachyon condensation the   Equation of State  changes 
from $w=0$ (tachyon away from its peak)  to $w=-1$ 
(as the tachyon climbs up its potential peak), 
at the same time the underlying  universe goes from  a
contraction phase to a deflation.

\begin{figure}
\centering
\includegraphics[width= 0.8\textwidth]{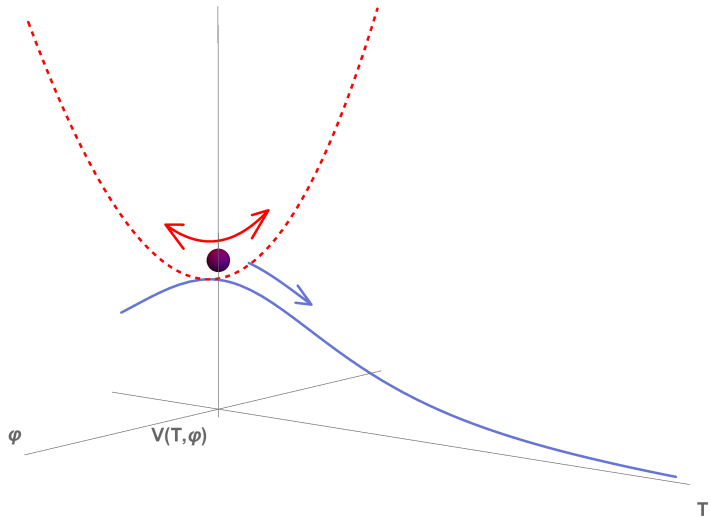}
\caption{A two-field potential can create a dynamical false vacuum  during the 
bounce phase. The blue solid line is the potential hill of tachyon field $T$ while 
the red dashed line corresponds to the potential valley governing  the Higgs field 
$\phi$ while at the peak of the tachyon potential.
When the Higgs  field is performing fast oscillation in its  potential valley, the tachyon field is dynamically locked at its  zero point.
The locked tachyon field behaves like ``dark energy''  sourcing  an exponential expansion (the locked inflation), which resembles an inflationary universe. 
As the universe expands, the amplitude of the scaler field is redshifted and 
D3-anti-D3-brane pairs annihilate to trigger the  tachyon condensation. 
The tachyon field finally rolls down the blue potential hill to ``reheat''  the universe.}
\label{fig_two_field_potential}
\end{figure}

\textbf{The Bounce Point: }
At the  bounce point of the CSTB universe,
it is the interplay of the curvature term and this vacuum energy that  brings  about 
a bounce  during which the underlying universe evolves  from $H < 0$ (contraction)  
to $H=0$ (bounce),  and then $H > 0$ (expansion) .
Furthermore the universe is dynamically stabilised at  a minimal radius, 
$$ 
  \frac{1}{a^{2}_{min} } 
  = \frac{8\pi}{3} G_{N} V_{0} 
  = \frac{4}{3} \frac{G_{N}}{g_{s}} (\frac{m_{s}}{2\pi})^{4}~,
  $$
when  $H=0$.  ${g_{s}} $ is the perturbative string coupling constant and ${m_{s}}$ is 
the string mass scale. 
As the tachyon reaches its potential peak it is  being dynamically locked at its potential  peak by the fast oscillations  of the Higgs field around its 
{\it minimum} at $\phi=0$. 
The  Friedmann  equation is  approximately the  same throughout with the Hubble 
parameter  goes from $H <0$ to  $H = 0$ and then to $H > 0$
corresponding to  the background universe  going  through a deflation, a bounce and 
then  a period of  inflation.  
Fig.~\ref{fig_two_field_potential} depicts the two-field
potential and how the dynamical false vacuum takes shape.
 A locked inflation ensues in the false vacuum.

\textbf{The cosmic evolution after the bounce: } 
At the peak of the tachyon potential the universe undergoes a ``locked inflation'' driven  by the tachyon's vacuum energy.   
%($w_{locked inflation}=-1$). 
  As the locked inflation proceeds,
the amplitude of $\phi$ oscillations  are red-shifted.
As  the Higgs oscillation energy   ceases  to dominate the energy density,
the tachyon is released from its potential peak and  it undergoes  a tachyon 
condensation while  the background universe is  expanding  in power law.  
 The tachyon behaves like a cold  
 matter~\cite{Sen:2002nu,Sen:2002in,Sen:2002an,Sen:2004nf}.
In other words, the equation of state (EoS) changes from $w=-1$ to
$w=0$ during the tachyon condensation~\footnote{%
During the period of rolling inflation and tachyon condensation
ordinary matter is generated and this may change the EoS of the underlying universe to be radiation dominated prior of matter domination.  
But this is outside the scope of this paper.  We will study  in detail this matter generation process in a forthcoming publication.
}.
After the tachyon condenses, the  universe goes onto a decelerating expansion stage
driven by the tachyon matter ($w=0$), which  evolves eventually  into our
present universe.

\subsection{Tensorial Perturbations}
Different from the inflationary scenario, the Coupled-Scalar-Tachyon bounce (CSTB) 
universe model  generates perturbations during the pre-bounce contracting phase.
In the  cosmological  perturbations  theory, the metric perturbations
are  sourced by  the perturbations of the background matter fields.
 To start the analysis, the  classical Lagrangian of CSTB  model is  expanded up to 
 second order in perturbations,  $\delta T$ and $\delta \phi$,
\begin{equation}  \label{eq:perturbed-action}
S=\int \sqrt{-g} \, d^4x \, \Big(  \frac{R}{16\pi G}
    + \mathscr{L}(T+\delta\,T,\phi+\delta\phi) \Big),
\end{equation}
where $ R$ and $ g$, being  the Ricci scalar and the determinant of
the metric respectively, should  also be  expanded to second order
 in metric perturbations: %$\delta g_{\mu\nu}$:
\begin{equation}    
\label{perturbed-metric}
g_{\mu\nu}=g^{(0)}_{\mu\nu}+\delta g_{\mu\nu}~.
\end{equation}
Note that  $g^{(0)}_{\mu\nu}$, $T^{(0)}$, $\phi^{(0)}$ and $a$
are governed by the classical equations of motion.
To study the perturbations we take background spacetime to be~\cite{Mukhanov:1990me, mukhanov2005physical},
$$
g_{\mu\nu}^{(0)} = diag\{-1,a(t)^2,a(t)^2,a(t)^2\}.
$$
There are two  independent spin-2   tensorial modes of perturbations  
in the $ \hat{k}=\hat{z}$ direction: 
\begin{equation}  
 \label{eq:perturbed-metric1}
\delta g_{\mu\nu} \equiv  h_{\mu\nu} =
a(t)^2  \left(\begin{array}{lcl}
         \begin{aligned}
          & 0 \quad & 0 \qquad 		   &  0  \qquad   &  0 \\
          & 0 \quad & h_{\times}\qquad &  h_+ \qquad  &  0  \\
          & 0 \quad & h_{+} \qquad	   & -h_{\times}  &  0  \\
          & 0 \quad & 0 \qquad         &   0  \qquad  &  0
         \end{aligned}
\end{array}\right)~.
\end{equation}

With  the perturbed action (Eq.~\ref{eq:perturbed-action}) 
one obtains the equations of motion for the two independent modes of metric tensors
perturbations,  $h_{+}$ and $h_{\times}$:
\begin{equation}  \label{eq:hx} %\nonumber
\begin{aligned}
 &
\frac{a^2}{8\pi{G}}\, (\ddot{h}_\times + 3H\dot{h}_\times -
                       a^{-2}\,\nabla^2  h_\times)  \\
= &
{h_\times}(\phi_{,1}^2+\phi_{,2}^2)
- 2 (\phi_{,1}\delta\phi_{,1} - \phi_{,2}\delta\phi_{,2})
+
\frac{V_0tanh(\frac{T}{\sqrt{2}})(T_{,1}^2-T_{,2}^2)\delta{T}}%
      {\sqrt{2} cosh(\frac{T}{\sqrt{2}})
    {\sqrt{1+g_{(0)}^{\mu\nu}\partial_\mu{T}\partial_\nu{T}}}
       }
\\
&+{h_\times} \frac{V_0(T_{,1}^2+T_{,2}^2)}
{cosh(\frac{T}{\sqrt{2}})\sqrt{1+g_{(0)}^{\mu\nu}
\partial_\mu{T} \partial_\nu{T}} }
-\frac{2 V_0 (T_{,1}{T}\delta T_{,1}-T_{,2} \delta T_{,2})}
{cosh(\frac{T}{\sqrt{2}})
  \sqrt{1+g_{(0)}^{\mu\nu}\partial_\mu{T}\partial_\nu{T}}}  \\
& + \frac{V_0 ({T_{,1}^2-T_{,2}^2})}{2cosh(\frac{T}{\sqrt{2}})
(1+g_{(0)}^{\mu\nu}\partial_\mu T\partial_\nu T)^{3/2}}
\Big[ g_{(0)}^{\mu\nu}\partial_\mu\,T\partial_\nu\delta{T}
   -  {a^{-2}}{h_+} T_{,1}T_{,2}
    - {2a^{-2}} h_\times (T_{,1}^2-T_{,2}^2)
    \Big]~,
\end{aligned}
\end{equation}
and
\begin{equation}  \label{eq:h+}
\begin{aligned}
& \frac{a^2}{8\pi{G}}
   (\ddot{h}_+   +  3H \dot{h}_+   - a^{-2}\nabla^2\, h_+)\\
= &  h_\times (\phi_{,1}^2+\phi_{,2}^2)   \,
    -2(\phi_{,1}\delta\phi_{,2}+\phi_{,2}\delta\phi_{,1})
    + \frac{\sqrt{2}\,V_0\,tanh(\frac{T}{\sqrt{2}})\,T_{,1}\,T_{,2}\,\delta{T}}{cosh(\frac{T}{\sqrt 2})\sqrt{1+g_{(0)}^{\mu\nu}\partial_\mu{T}\partial_\nu{T}}}\\
& + h_\times \frac{V_0 (T_{,1}^2 + T_{,2}^2)}
{cosh(\frac{T}{\sqrt{2}})
    \sqrt{1+g_{(0)}^{\mu\nu}\partial_\mu{T}\partial_\nu{T}}}
-\frac{2 V_0 (T_{,1}\delta T_{,2}+T_{,2}\delta T_{,1})}
{cosh(\frac{T}{\sqrt{2}})\sqrt{1+g_{(0)}^{\mu\nu} \partial_\mu\,T\partial_\nu{T}}}  \\
& + \frac{V_0 T_{,1}T_{,2}}
     {2cosh(\frac{T}{\sqrt{2}})
     (1+g_{(0)}^{\mu\nu}\partial_\mu\,T\partial_\nu\,T)^{3/2}}
  \Big[ g_{(0)}^{\mu\nu}\partial_\mu\,T\partial_\nu \, \delta{T}
        -{a^{-2}}h_+ T_{,1}T_{,2}- 2a^{-2} h_\times(T_{,1}^2-T_{,2}^2)
      \Big]~,
\end{aligned}
\end{equation}
where we denote  $ \partial_i T$  and $ \partial_i \phi$ by
$ T_{,i}$ and $ \phi_{,i}$  ($ i=1,2,3$).

The equations of motion~(Eq.~\ref{eq:hx}) and~(Eq.~\ref{eq:h+})  show 
that the spatial derivatives of the perturbations always couple to  the
spatial derivatives of the background fields to first order.
Under the assumption of the Cosmological Principle that the Universe is homogeneous
and isotropic at large scales,  the spatial derivative of the
background fields vanishes. 
The equations of motion~(Eq.~\ref{eq:hx}) and~(Eq.~\ref{eq:h+}) thus reduce to
\begin{equation}   \label{eq:tensor-modes}
\ddot{h}_{\times,+}  +   3H\dot{h}_{\times,+}
-a^{-2}{\nabla^2 \, h_{\times,+}} =0.
\end{equation}
Applying Fourier transformation to~(Eq.~\ref{eq:tensor-modes})  and 
going to  conformal time, $\eta$,  we obtain,
\begin{equation}     \label{eq:tensor-k-eta}
h_{k}^{\prime\prime}+2\frac{a^\prime}{a}h_{k}^\prime+k^2h=0,
\end{equation}
an equation obeyed by each  Fourier mode, $h_{k}$.
 $ ^\prime$ denotes  a derivative with respect to
the conformal time and  $ h_{k}=h_{k}(\eta,k)$. % is   a function  of  .

\textbf{Discussion:}  One  can indeed  check that the perturbation terms
of the tachyon and scalar fields  sourcing  the tensor modes
originate  from  $  g_{(0)}^{\mu\nu}\partial_\mu T\partial_\nu T$ and
$  g_{(0)}^{\mu\nu}\partial_\mu\phi\partial_\nu\phi$ in  the perturbed
Lagrangian~(Eq.~\ref{eq:perturbed-action}).
Only those terms that are  multiplied by the tensor perturbations $h_+$ and
$h_{\times}$  in the perturbed Lagrangian remain, upon
variations   with respect to $h_+$ and $h_{\times}$.
As a consequence, the  spatial derivatives of the  perturbations  coupled  to  the
spatial derivatives of the background fields  can eventually remain in the equations
of motion for  the tensor modes.
 The time derivatives of  $ h$  and those of $ \delta T$ and $ \delta\phi$
 never have a chance to couple with the tensor perturbations of   $h$ to the
quadric order.  This is simply because the tensor perturbations are the metric
perturbations  in  the purely spatial part.
This confirms an important observation:  tensorial perturbations  of a homogeneous  
and isotropic universe cannot be sourced by the background fields and their  fluctuations~\cite{mukhanov2005physical, Mukhanov:1990me, Deruelle:1995kd}.

\section{The  primordial tensor  perturbations in a simplified bounce}
\label{sec:simplified_bounce}

The evolution of the background universe  is governed by the classical equations of motion.
In  the context of cosmic perturbations theory  the known array of ``matter'' content
is seeded by the small/quantum fluctuations of fields above this classical background.
Care should be exercised to ensure that these primordial ``matter'' density
perturbations remain small perturbations at all time; otherwise these matter
constituents would have contributed significantly in the cosmic budget and
influence the  evolution of the underlying cosmic background.  
In~\cite{Li:2013bha, Li:2012vi} the spectrum of primordial density  perturbations 
of the  scalar modes  are obtained.  Care has been taken to compute the implicit time
 dependence  in the k-modes as they exit their effective horizon at different times. 
The final spectrum of primordial density perturbations is shown to be scale 
invariant as well as stable against time evolution~\cite{Li:2013bha, Ming:2017dtm}.

Likewise the dynamics of the tensorial perturbations in the metric can be studied 
in an analogous manner. 
For the current analysis of the tensorial modes in metric perturbations,
we write down the analytic  expressions  of the  scale factor using the EoS  at   each  cosmic epoch:
\begin{equation}
\label{eq:a-bounce}
 a(t)=\left\{\begin{array}{lcl}
\begin{aligned}
&a_{ec}(1+\frac{3H_{ec}}{2}(-t-t_{ec}))^{\frac{2}{3}},\  &t<-t_{ec}\\
&a_{*} \frac{e^{-H_{ec} t}+e^{H_{ec}  t}}{2},\        &-t_{ec}<t<t_{ec}\\
&a_{ec}(1+\frac{3H_{ec} }{2}(t-t_{ec}))^{\frac{2}{3}},\  &t>t_{ec}
\end{aligned}
\end{array}\right.
\end{equation}
with the subscript 'ec' corresponding to the end of the reverse tachyon condensation~\cite{Sen:2002in}. 
 This corresponds to  the end of the  matter-dominated contraction epoch in  
the CSTB universe,  after which  the tachyon  will get up to the peak of its 
potential hill and trigger a period of  locked inflation.
%% t_{ec}
We shall henceforth denote the Hubble parameter at this point by 
\begin{equation}
\label{eq:Hec} 
H_{ec} \sim -\sqrt{\frac{8\pi V_0}{3M_p^2}}~,  
\end{equation}
with the corresponding scale factor denoted by $a_{ec}$ and the physical time denoted by  $t_{ec}$. 
%The minimal radius of the universe attained at the bounce point is given by:
%\begin{equation} \label{eq:amin}
%  \frac{1}{a^{2}_{*} } 
%  = \frac{8\pi}{3} G_{N} V_{0} 
%  = \frac{4}{3} \frac{G_{N}}{g_{string}} (\frac{m_{string}}{2\pi})^{4}~.
%\end{equation}

\begin{figure}[htbp]
\centering
\includegraphics[width=0.8\textwidth]{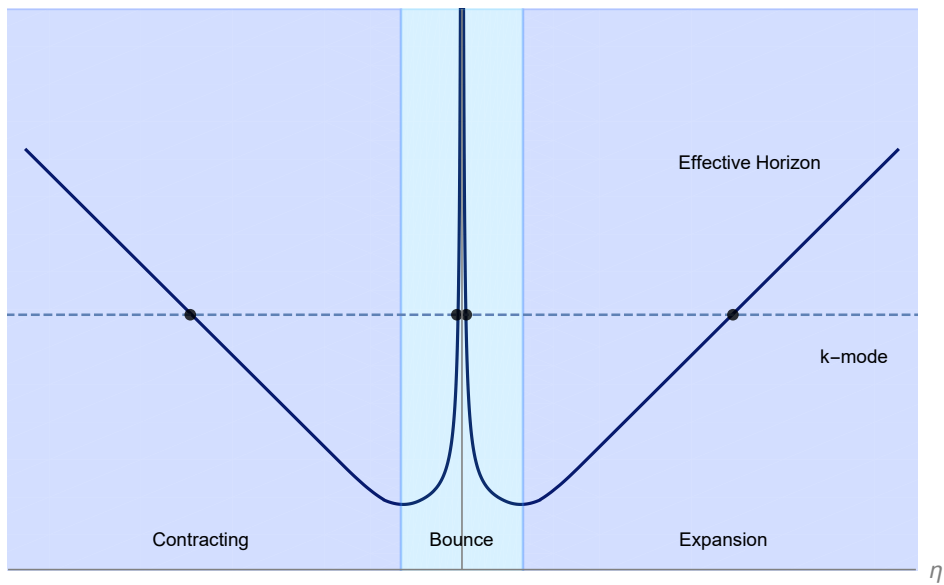}
\caption{The evolution of the effective horizon in  conformal time, 
in which the k-modes are the ``co-moving'' k-modes and do not  change with
the conformal time. It is thus  obviously that one k-mode crosses
the effective horizon  four times (the black dots on the diagram) in a bounce universe.
We will henceforth simplify the bounce phase (the light blue region) by  a point;
only the exit and reentry during the tachyon dominated contracting and expansion phases
 (the dark blue region) are considered when solving perturbations equations~(Eq.~\ref{eq:a_eta_simple}).  Analysis of the scalar perturbations
 in~\cite{Li:2013bha,  Li:2012vi}
 lends the  theoretical support to the said idealisation.}
\label{fig_horizon}
\end{figure}

\subsection{The horizon crossing of tensor modes}
\label{sec:horizon_cross}
From the dynamics of the scale factor in each epoch one can introduce the 
effective horizon as seen by each Fourier mode by taking the  inverse of the Hubble parameter,
\begin{equation}
\label{effec_horizon_conformal}
\frac{1}{|\mathcal{H}|}=\frac{1}{a}\frac{1}{|H|}=\frac{a}{|a'|}.
\end{equation}
Note here we have made no prior assumption of the Hubble parameter being 
a constant, as it is usually assumed in the case of inflation scenario.
The effective horizon is simply the Hubble radius in  conformal time, as 
shown in  Fig.~\ref{fig_horizon}.

The horizon crossing conditions require that  the wavelength of a given k-mode be equal to  its  effective Hubble horizon:
\begin{equation}
\label{horizon_crossing_condition2}
k=a|H|.
\end{equation}
The scale factor evolves as a power law  in conformal time during the tachyon matter domination,
\begin{equation}
a(\eta)=\eta^{\nu}, 
\end{equation}
%The  horizon crossing conditions, given by~(Eq.~\ref{horizon_crossing_condition2}), 
%is thus  re-expressed as
leading to 
\begin{equation}
\label{horizon_crossing_condition3}
k|\eta|=|\nu|~.
\end{equation}
One can see from  Fig.~\ref{fig_horizon} that long wavelength modes exit  
the horizon earlier and reenter  later than short wavelength ones.
As the conformal time approaching  zero in~(Eq.~\ref{horizon_crossing_condition3}),
the inequality $k|\eta|<|\nu|$ implies that the wavelength of the k-mode is larger 
than the effective horizon. On the other hand, the large time limit
$\eta\rightarrow\infty$ corresponds to the epochs when the k-modes are well inside the horizon. 
In summary,
\begin{equation}
\label{eq:k_eta_simplified_bounce}
\left\{\begin{array}{lcl}
\begin{aligned}
&k\eta\rightarrow -\infty,\ & \mathrm{in\,  the\,  far\, past; }\\
&k\eta\rightarrow 0,      \ &\mathrm{outside\, the\, horizon;} \\
&k\eta\rightarrow +\infty,\ & \mathrm{in \, the \, far\, future.}\\
\end{aligned}
\end{array}\right.
\end{equation}
Fig.~\ref{fig_horizon} corroborates with~(Eq.~\ref{eq:k_eta_simplified_bounce}) above.

\subsection{Idealisation of the bounce process } 

Since the coupled tachyon condensation and the reversal of tachyon condensation~\cite{Sen:2002nu} persist a short time relative to the 
contraction and expansion phases of CSTB universe,  we could have set them to  two points
 on  the time axis when studying the cosmic  evolution~\cite{Li:2011nj, Cheung:2016oab}.
This is especially the case when  studying dynamics of the primordial perturbations of matter~\cite{Li:2013bha}.  Furthermore we shall ignore the evolution at the bounce point by treating the bounce phase as a point. This is justified by the fact that we are interested 
in the large scale structures  resulting  from the gravitational waves: the long-wavelength 
modes are not sensitive to the small scale fluctuations in the cosmic background as they 
have long decoupled from the dynamics (out of the horizon at a much earlier time). 

The evolution of the CSTB universe is then turned  into a two-phase evolution: the 
pre-bounce contraction and the post-bounce expansion, with 
 the deflation, the locked inflation, and the inflation be collectively represented by 
 ``the bounce point.''
As a result, the scale factor of a bounce universe is given by
\begin{equation}
\label{eq:a_eta_simple}
a(\eta)=\left\{\begin{array}{lcl}
\begin{aligned}
&a_{ec} (1-\frac{H_{ec} a_{ec}\eta}{2})^{2},\  &\eta<0\\
&a_{ec} (1+\frac{H_{ec} a_{ec}\eta}{2})^{2},\ &\eta\ge0\ ,
\end{aligned}
\end{array}\right.
\end{equation}
where $a_{ec}$ is the scale factor at which the end of the tachyon-matter-dominated 
contration phase.  Note that  $a_{min}$  
is the smallest  scale  the CSTB universe can possibly  attain  at the  ``bounce point''  given by
\begin{equation}       %% a_{min}
         \label{eq:amin}
 a_{min} % = \large(\frac{3}{4} g_{string}\large)^{\frac{1}{2}} \frac{l_{string}^2}{l_{Planck}},
= \large(g_{s}\large)^{\frac{1}{2}} \frac{l_{s}^2}{l_{p}}~, 
 \end{equation}
which is lower than the Planck scale, where $g_{s}$ is the perturbative string coupling,  
$l_{s}$ is the string length and $l_{pl}$  is the Planck length.

\subsection{Solving the tensor mode equations}
With the equations of motion for the Fourier modes  of the tensor 
perturbations~(Eq.~\ref{eq:tensor-k-eta}) 
\begin{equation}
\label{eq:k-mode}
h^{\prime\prime}+ 2\frac{a^\prime}{a}h^\prime+k^2h=0~,
\end{equation}
is given in  Section~\ref{sec:tensor-modes} above, 
one applies the usual change of variables, $u=ah$,  and arrives at the familiar equation,
\begin{equation}
\label{eq_u}
u^{\prime\prime}+(k^2-\frac{a''}{a})u=0~.
\end{equation}

In the  matter-dominated phases of the CSTB model, the scale factor evolves by power law with  $\nu=2$. The solution of~(Eq.~\ref{eq_u}) takes  a  general form
\begin{equation}
\label{eq:solution_u_general}
u=C_k \eta^{\frac{1}{2}}J_{\frac{3}{2}}(k\eta)+D_k \eta^{\frac{1}{2}}J_{-\frac{3}{2}}(k\eta)~.
\end{equation}
%during the  contraction and expansion phases.  
Whereas the coefficients $C_k$ and $D_k$ %in~(Eq.~\ref{eq:solution_u_general}) 
are constants in $\eta$ they  can be
functions in $k$, to be determined by initial conditions and  matching conditions. Thus 
\begin{equation}
\label{eq:h_solution_simple}
h=\left\{\begin{array}{lcl}
\begin{aligned}
&\frac{C_{ctr}}{a(\eta)}\Big(\frac{2}{H_{ec}a_{ec}}-\eta\Big)^{\frac{1}{2}}J_{\frac{3}{2}}(\frac{2k}{H_{ec}a_{ec}}-k\eta)+
\frac{D_{ctr}}{a(\eta)}\Big(\frac{2}{H_{ec}a_{ec}}-\eta\Big)^{\frac{1}{2}}J_{-\frac{3}{2}}(\frac{2k}{H_{ec}a_{ec}}-k\eta),\ \eta<0\\
&\frac{C_{exp}}{a(\eta)}\Big(\frac{2}{H_{ec}a_{ec}}+\eta\Big)^{\frac{1}{2}}J_{\frac{3}{2}}(\frac{2k}{H_{ec}a_{ec}}+k\eta)+
\frac{D_{exp}}{a(\eta)}\Big(\frac{2}{H_{ec}a_{ec}}+\eta\Big)^{\frac{1}{2}}J_{-\frac{3}{2}}(\frac{2k}{H_{ec}a_{ec}}+k\eta),\ \eta\ge0,
\end{aligned}
\end{array}\right.
\end{equation}
 are the most general solutions to~(Eq.~\ref{eq:k-mode}), with $J_{j}$ being 
  the Bessel functions in the j-th order. 
% The $J_{\frac{3}{2}}$ and $J_{-\frac{3}{2}}$  are the $\frac{3}{2}$-order and
% $(-\frac{3}{2})$-order Bessel functions respectively.
The two  branches of solutions in~(Eq.~\ref{eq:h_solution_simple})
correspond to the contraction ($\eta<0$)  and the expansion ($\eta> 0$) phases.
They are to be connected by the matching conditions at the bounce point when $\eta = 0$.
Upon obtaining the complete  wave function of gravitational waves in the bounce universe
  one can then  study the properties of the primordial gravitational waves spectrum and their observational signatures.

\subsection{The short wavelength and long wavelength limits}
Applying the limits~(Eq.~\ref{eq:k_eta_simplified_bounce})
 to the  general solutions~(Eq.~\ref{eq:h_solution_simple}),
 we obtain the expression for each k-mode in one of the following situation:
\begin{itemize}
\item
During contraction~($\eta<0$), the k-mode is well inside the
horizon~($\eta\rightarrow-\infty$):
\begin{equation}
\label{h_simple_contracting_inside}
h=%\left\{\begin{array}{lcl}
%\begin{aligned}
-\frac{C_{ctr}}{a(\eta)}\sqrt{\frac{2}{\pi k}}cos(\frac{2k}{H_{ec} a_{ec}}-k\eta)-\frac{D_{ctr}}{a(\eta)}\sqrt{\frac{2}{\pi k}}sin(\frac{2k}{H_{ec}a_{ec}}-k\eta)~.%\ &\eta\rightarrow -\infty\\
%&\frac{1}{3}\sqrt{\frac{2}{\pi}}{\frac{C_{ctr}}{a(\eta)}}\Big(\frac{2}{H_{ec} a_{ec}}-\eta\Big)^2k^{\frac{3}{2}}-\sqrt{\frac{2}{\pi}}\frac{D_{ctr}}{a(\eta)}\Big(\frac{2}{H_{ec} a_{ec}}-\eta\Big)^{-1} k^{-\frac{3}{2}},\ &\eta\rightarrow 0
%\end{aligned}
%\end{array}\right.
\end{equation}
\item
During contraction~($\eta<0$), the k-mode is far outside the
horizon~($\eta\rightarrow 0^{-}$):
\begin{equation}
\label{h_simple_contracting_outside}
h=%\left\{\begin{array}{lcl}
%\begin{aligned}
%&-\frac{C_{ctr}}{a(\eta)}\sqrt{\frac{2}{\pi k}}cos(\frac{2k}{H_{ec} a_{ec}}-k\eta)-\frac{D_{ctr}}{a(\eta)}\sqrt{\frac{2}{\pi k}}sin(\frac{2k}{H_{ec}a_{ec}}-k\eta),\ &\eta\rightarrow -\infty\\
%&
\frac{1}{3}\sqrt{\frac{2}{\pi}}{\frac{C_{ctr}}{a(\eta)}}\Big(\frac{2}{H_{ec} a_{ec}}-\eta\Big)^2k^{\frac{3}{2}}-\sqrt{\frac{2}{\pi}}\frac{D_{ctr}}{a(\eta)}\Big(\frac{2}{H_{ec} a_{ec}}-\eta\Big)^{-1} k^{-\frac{3}{2}}~.%\ &\eta\rightarrow 0
%\end{aligned}
%\end{array}\right.
\end{equation}
\item
During expansion~($\eta> 0$), the k-mode is far outside  the
horizon~($\eta\rightarrow 0^{+}$):
\begin{equation}
\label{h_simple_expansion_outside}
h=%\left\{\begin{array}{lcl}
%\begin{aligned}
%&
\frac{1}{3}\sqrt{\frac{2}{\pi}}{\frac{C_{exp}}{a(\eta)}}\Big(\frac{2}{H_{ec} a_{ec}}+\eta\Big)^2k^{\frac{3}{2}}-\sqrt{\frac{2}{\pi}}\frac{D_{exp}}{a(\eta)}\Big(\frac{2}{H_{ec} a_{ec}}+\eta\Big)^{-1} k^{-\frac{3}{2}};%\ &\eta\rightarrow 0\\
%&-\frac{C_{exp}}{a(\eta)}\sqrt{\frac{2}{\pi k}}cos(\frac{2k}{H_{ec}a_{ec}}+k\eta)-\frac{D_{exp}}{a(\eta)}\sqrt{\frac{2}{\pi k}}sin(\frac{2k}{H_{ec}a_{ec}}+k\eta),\ &\eta\rightarrow +\infty
%\end{aligned}
%\end{array}\right.
\end{equation}
\item
During expansion~($\eta> 0$), the k-mode is well inside  the
horizon~($\eta\rightarrow +\infty $):
\begin{equation}
\label{h_simple_expansion_inside}
h=-\frac{C_{exp}}{a(\eta)}\sqrt{\frac{2}{\pi k}}cos(\frac{2k}{H_{ec}a_{ec}}+k\eta)-\frac{D_{exp}}{a(\eta)}\sqrt{\frac{2}{\pi k}}sin(\frac{2k}{H_{ec}a_{ec}}+k\eta)~.
\end{equation}
\end{itemize}
The scale factor, $a(\eta)$, is given by~(Eq.~\ref{eq:a_eta_simple}) in each of the corresponding phases.

\paragraph{Bunch-Davies vacuum as an  initial condition:}

In the far past  we  expect the universe would have relaxed into its ground state and 
a natural choice is the Bunch-Davies vacuum
\begin{equation}
\label{eq:BD_vacuum}
h \sim \frac{1}{m_{pl}}\frac{1}{\sqrt{2k}}\frac{e^{-ik\eta}}{a(\eta)},
\end{equation}
as an initial condition for  the primordial gravitational waves at a time far 
prior  to the bounce  when the modes are well inside the horizon. 
Matching the first branch of the 
solution(Eq.~\ref{h_simple_contracting_inside}) to the Bunch-Davies 
vacuum(Eq.~\ref{eq:BD_vacuum}), we find the coefficients during the contracting phase:
\begin{equation}
\label{initial_condition_simplified_bounce}
\left\{\begin{array}{lcl}
\begin{aligned}
&C_{ctr}=-\frac{\sqrt{\pi}}{2m_{pl}}~,\\
&D_{ctr}=-i\frac{\sqrt{\pi}}{2m_{pl}}~.
\end{aligned}
\end{array}\right.
\end{equation}
Given the functional forms  of tensorial perturbations, once  we  find the right 
combinations of coefficients  $(C , D)$ to capture the physics of interest  
at the bounce point, the entire evolution of gravitational waves in the CST 
bounce universe is then fully  determined. In particular we need to find  
the rules to connect coefficients $(C_{exp},D_{exp})$ to $(C_{ctr},D_{ctr})$
 as the expansion phase meets with the contraction phase at $\eta=0$. 
In the following  we  suggest  two reasonable matching conditions to capture the 
essential physics of our interest.

\begin{figure}
\centering
\subfigure[]
{%
\label{fig:subfig1:a} %% label for first subfigure
\includegraphics[width=0.48\linewidth]{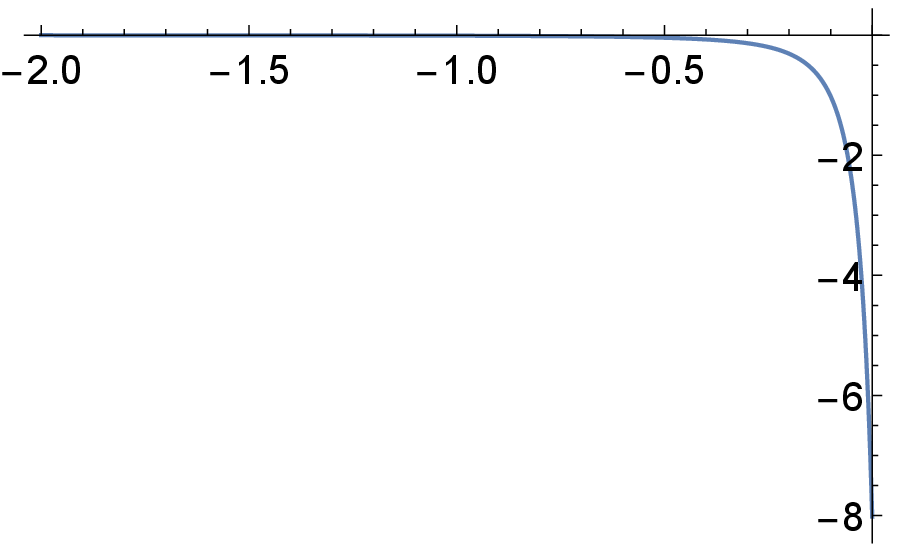}}
\subfigure[]
{%
\label{fig:subfig1:b} %% label for second subfigure
\includegraphics[width=0.48\linewidth]{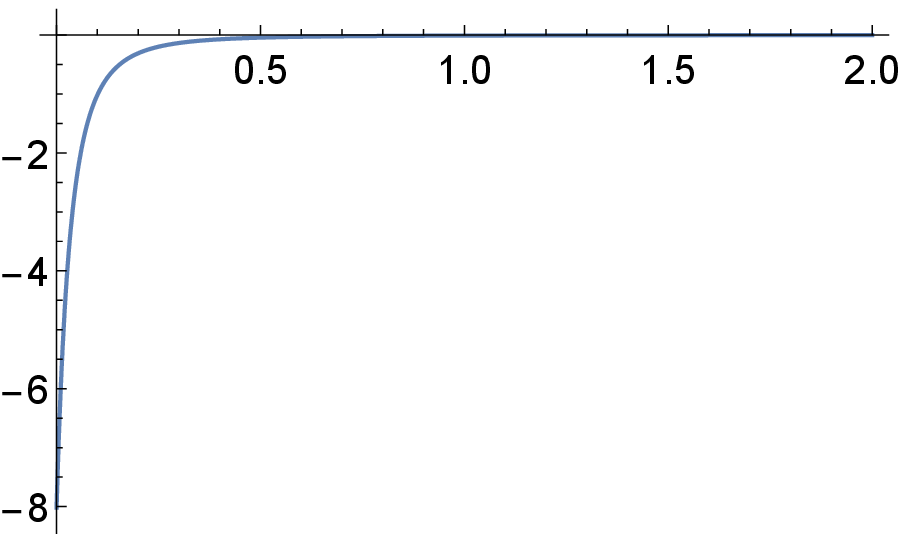}}
\subfigure[]
{%
\label{fig:subfig1:c} %% label for first subfigure
\includegraphics[width=0.45\linewidth]{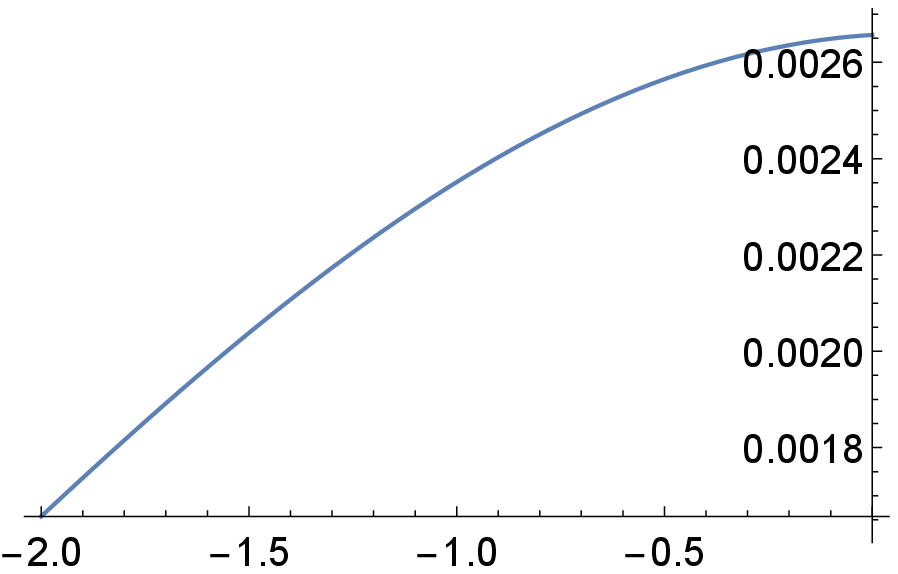}}
\subfigure[]
{%
\label{fig:subfig1:d} %% label for second subfigure
\includegraphics[width=0.50\linewidth]{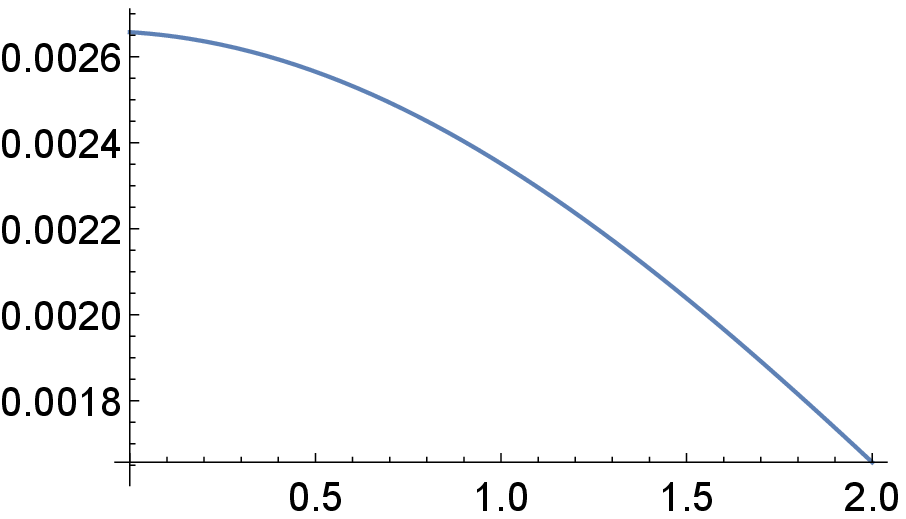}}
\caption{Matching the tensor modes across the bounce: Two independent solutions to 
the gravitational wave equations  are plotted with $H_{ec}=20$, $a_{ec}=1$ and $k=1$. 
 The  gravitational waves functions  before the bounce are plotted in 
 Fig.~\ref{fig:subfig1:a} and Fig.~\ref{fig:subfig1:c} on the left; 
 whereas  the gravitational waves functions  after the bounce are depicted   
  in  Fig.~\ref{fig:subfig1:b} and Fig.~\ref{fig:subfig1:d}  on the right.
Our  task is to match the two wave  functions on the right  to a combination 
of wave functions on the left by relating $(C_{ctr},D_{ctr})$ to $(C_{exp},D_{exp})$ 
at  $\eta=0$.}
\label{fig:subfig1} %% label for continuous matching 
\end{figure}

\subsection{Matching conditions}
%\paragraph{The problem of matching:}
To state the problems we are facing, 
we need to match the two solutions~(Eq.\ref{eq:h_solution_simple}) 
to the gravitational waves equation  at the bounce point, where the 
solution~(Eq.\ref{h_simple_contracting_outside})  in the contracting  phase 
meets with the solution (Eq.\ref{h_simple_expansion_outside}) in the 
expansion phase, when the conformal time approaches zero.
This is visualised in Fig~\ref{fig:subfig1} and  Fig.~\ref{fig:subfig}  
%Fig~\ref{fig:subfig1} and  Fig.~\ref{fig:subfig}  
numerically with $H_{ec}=20$, $a_{ec}=1$ and $k=1$.

\subsubsection*{I. The continuous matching condition}
The first matching condition we propose is the conventional one: to ensure the continuity of the wave functions across the bounce by 
demanding  the wave functions and their derivatives  be continuous throughout 
the bounce phase following~\cite{Deruelle:1995kd}. 
\begin{equation}
\label{eq:matching_conditions_simplified_bounce}
\left\{\begin{array}{lcl}
\begin{aligned}
&C_{exp}=C_{ctr}-6D_{ctr}(\frac{H_{ec} a_{ec}}{2k})^3,\\
&D_{exp}=-D_{ctr}.
\end{aligned}
\end{array}\right.
\end{equation}
The $D_{ctr}(\frac{H_{ec} a_{ec}}{2k})^3$ term in the matching condition 
(Eq.~\ref{eq:matching_conditions_simplified_bounce}) is much larger than $C_{ctr}$ 
and thus dominating $C_{exp}$. 
%The result is presented  by  numerics  in Fig.~\ref{fig:subfig1}.

\subsubsection*{Remarks on matching conditions}
The  seemingly complicated physical process of  ``bounce''  does allow 
a simple handling when gravitational waves spectrum calculations are concerned, as done in Section~\ref{sec:simplified_bounce}. 
The spectrum of primordial gravitational waves we are interested in lies in the 
low frequency (low energy) range. These modes of gravitational waves decouple much earlier 
long before  the scale of matter genesis sets in. 
Furthermore the minimal radius of the Coupled-Scalar-Tachyon bounce universe 
is  at/below the string scale~(\ref{eq:amin}).
Even though the physics may be  complicated at the bounce point, it does not 
affect the low-energy modes of gravitational waves  which  we are interested in.  
Thus we can simplify the bounce phase as we did in Section~\ref{sec:simplified_bounce}.
For our purpose we note that the  overall  continuity of GW wave-functions at 
the idealised  bounce point may break down. 
Fortunately there is another way to fix the coefficients across the bounce point.

\subsubsection*{II. A symmetric  matching condition}
\begin{figure}
\centering
\subfigure[]
{
\label{fig:subfig:a} %% label for first subfigure
\includegraphics[width=0.48\linewidth]{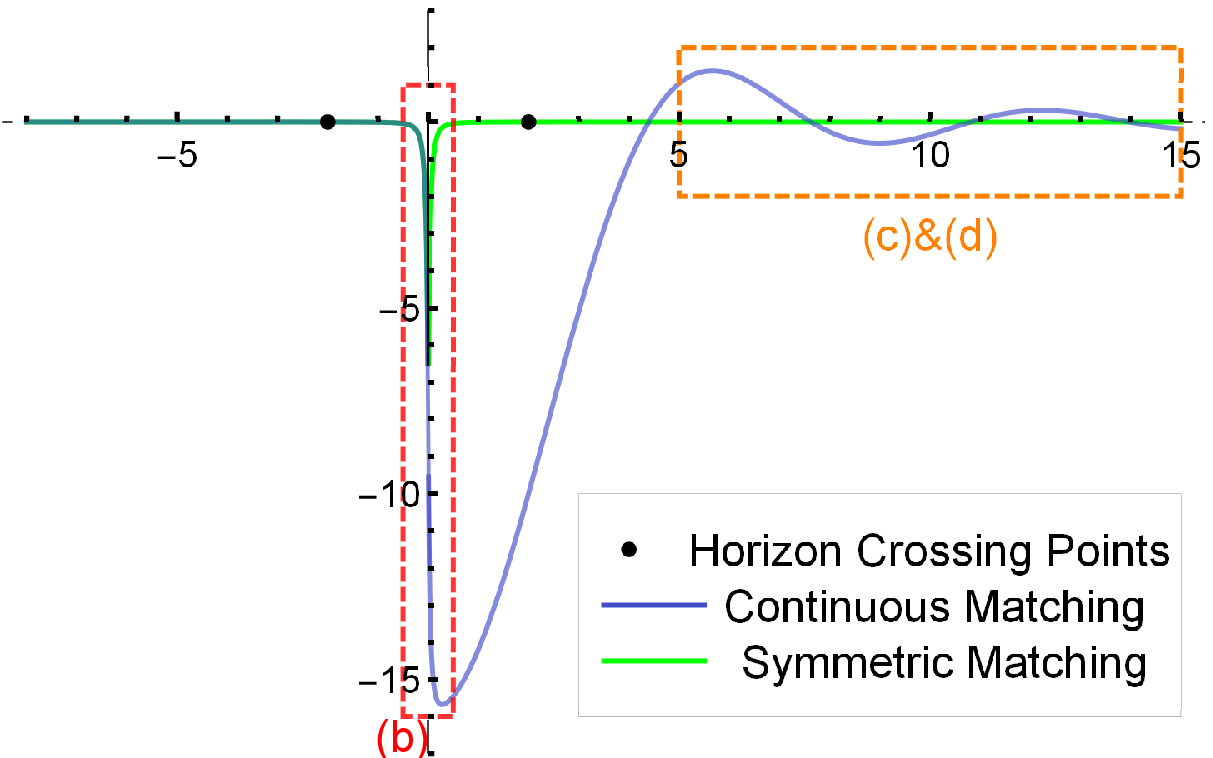}}
\subfigure[]
{
\label{fig:subfig:b} %% label for second subfigure
\includegraphics[width=0.48\linewidth]{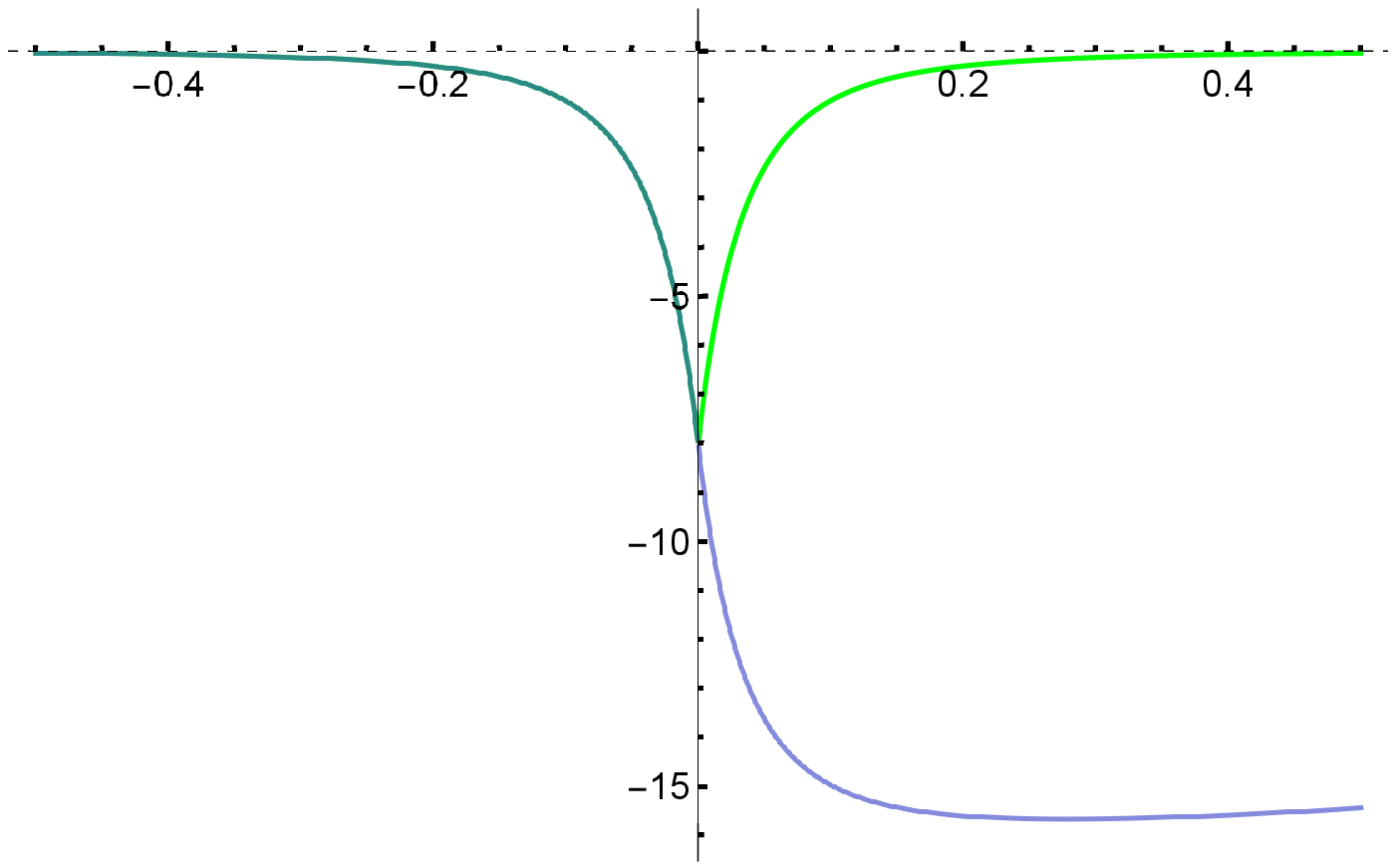}}
\subfigure[]
{
\label{fig:subfig:c} %% label for first subfigure
\includegraphics[width=0.45\linewidth]{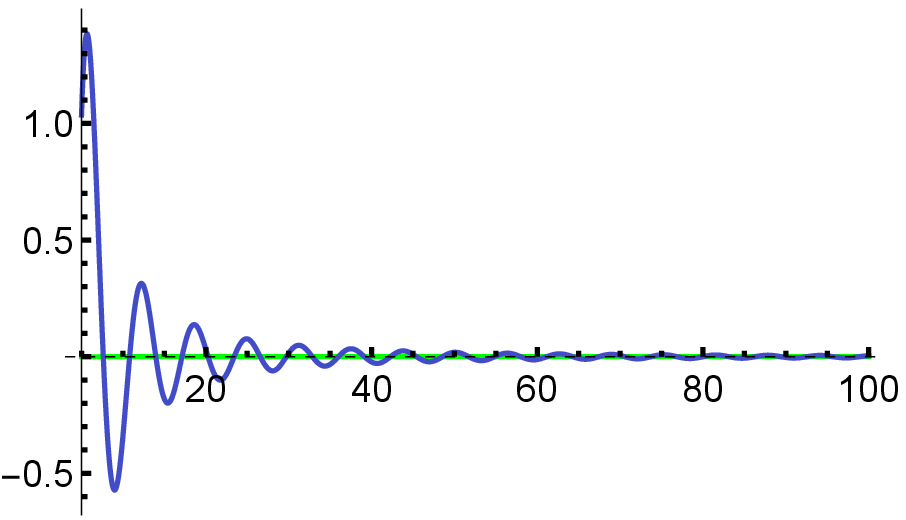}}
\subfigure[]
{
\label{fig:subfig:d} %% label for second subfigure
\includegraphics[width=0.50\linewidth]{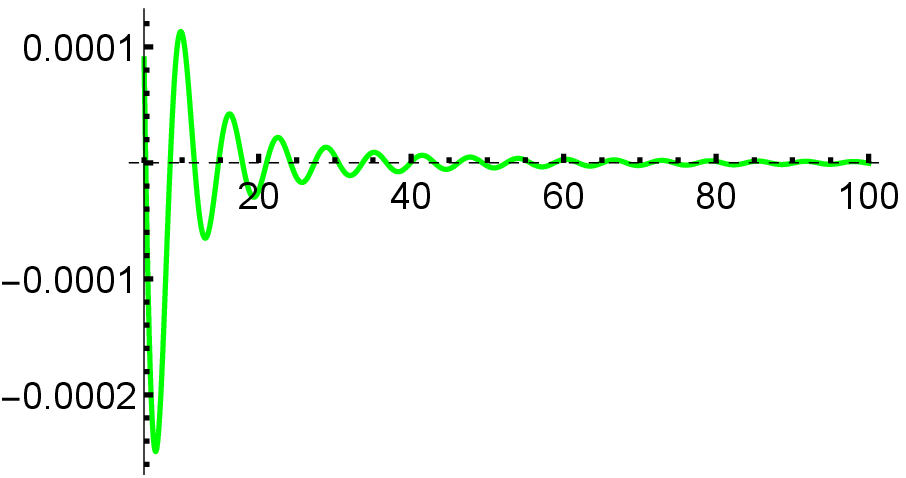}}
\caption{Two matching conditions yield different matched wave
functions after the bounce.
In subfigure~\ref{fig:subfig:a},  the blue curve corresponds to the 
GW wavefunction resulting  from the continuous matching while the green curve 
corresponds to the symmetrical matching condition.
Subfigures~\ref{fig:subfig:b} amplifies  the details of the wavefunctions 
at the vicinity of the bounce point. One sees  that  a smooth curve is 
obtained with the the continuous condition; while a symmetric wavefunction 
is obtained with the symmetric matching. 
After the bounce point, both matching conditions produce  oscillatory  modes. 
 Subfigures~\ref{fig:subfig:c} and~\ref{fig:subfig:d} show the scales 
 of these oscillations.
One also notes  that the continuous matching condition produces  a significantly 
larger amplitude  after the bounce than the symmetrical matching condition.}
\label{fig:subfig} %% label for entire figure 
\end{figure}

If we examine the scale factor before the bounce as well as after the bounce, 
we find that the derivative of the scale factor gets  a minus sign through 
a bounce point with the same expression. 
We thus  expect the derivative of  tensor modes changes by a minus sign 
after  the  bounce point. 
This allows us to  fix the coefficients after the bounce by
\begin{equation}
\label{eq:matching_symmetry_coefficients}
\left\{\begin{array}{lcl}
\begin{aligned}
&C_{exp}=C_{ctr},\\
&D_{exp}=D_{ctr}.
\end{aligned}
\end{array}\right.
\end{equation}
Combining the initial condition provided by the Bunch-Davis vacuum~(Eq.~\ref{initial_condition_simplified_bounce}) and the symmetric matching
condition~(Eq.~\ref{eq:matching_symmetry_coefficients}), we 
obtain the expressions   of ($C_{exp},D_{exp}$)
\begin{equation}
\label{eq:matching_symmetry_coefficients2}
\left\{\begin{array}{lcl}
\begin{aligned}
&C_{exp}=-\frac{\sqrt{\pi}}{2m_{pl}},\\
&D_{exp}=-i\frac{\sqrt{\pi}}{2m_{pl}}.
\end{aligned}
\end{array}\right.
\end{equation}
In Fig.~\ref{fig:subfig} the symmetrically matched wavefunctions are shown in green.

\subsubsection*{The power spectra resulting  from the two  matching conditions}
The  full analytical expressions for the  tensorial metric  perturbations
throughout the bounce can thus be obtained by combining the matched coefficients 
after the bounce given by~(Eq.~\ref{eq:matching_conditions_simplified_bounce}), or~(Eq.~\ref{eq:matching_symmetry_coefficients2}) as the case may be, with  
the coefficients before the bounce as provided by the Bunch-Davies
vacuum~(Eq.~\ref{initial_condition_simplified_bounce}).  
The full  wave-funtions are summarised in Table~\ref{Table_eta}  for 
 the  continuous gluing condition and %in Table~\ref{Table_eta} in  the case of 
the symmetric gluing condition.

\subsection{The time dependence in the power spectra}
The $\eta$-dependence of the gravitational wave functions  resulting  from both gluing conditions 
can be found in Table~\ref{Table_eta}. %and Table~\ref{Table_eta_symmetry}. 
We  see that the wave function  is indeed  symmetric in the case of the symmetric matching.
 While being outside of  the  horizon, the physical modes  are not frozen in  either  cases. 
 This is  a  remarkable   difference   from   inflation universe, as already  noted in~\cite{Wands:1998yp}. 
The primordial perturbations with the symmetrical matching have the same form as  the  BD vacuum 
while the continuous matching condition  produces  much  larger oscillations. 
Both matching conditions yield  the same $\eta$-dependence after the horizon reentry as the 
primordial perturbations decays in the same way as  the universe expands.
%%%% table of eta dependence ... 
\begin{table}[]\tiny
\centering
\begin{tabular}{|c|c|c|c|}
\hline
%\multirow{2}*{ }
\multicolumn{2}{|c|}{Continuous Matching}&\multicolumn{2}{|c|}{Symmetrical Matching}\\
%\cline{2-5}
\hline
~Analytical Solution
&$\eta^{\nu}$
&Analytical Solution
&$\eta^{\nu}$\\
\hline
%\begin{tabular}[c]{@{}c@{}}
%Before\\ horizon exit
%\end{tabular}
\multicolumn{4}{|l|}{Before horizon exit}\\
%\hline
$\displaystyle{\frac{1}{\sqrt{2k}}\frac{e^{-ik\eta}}{a(\eta)}}$
& $\displaystyle{(-\eta)^{-2}}$
&$\displaystyle{\frac{1}{\sqrt{2k}}\frac{e^{-ik\eta}}{a(\eta)}}$
&$\displaystyle{(-\eta)^{-2}}$\\
\hline
%\begin{tabular}[c]{@{}c@{}}
%From exit\\ to bounce
%\end{tabular}
\multicolumn{4}{|l|}{From exit to bounce}\\
%\hline
 $\displaystyle{-\frac{\sqrt{2}}{H_{ec}^2a_{ec}^3}(\frac{1}{3}-i(\frac{2}{H_{ec}a_{ec}}-\eta)^{-3}k^{-3})k^{\frac{3}{2}}}$
& $\displaystyle{(-\eta)^{-3}}$
&$\displaystyle{-\frac{\sqrt{2}}{H_{ec}^2a_{ec}^3}(\frac{1}{3}-i(\frac{2}{H_{ec}a_{ec}}-\eta)^{-3}k^{-3})k^{\frac{3}{2}}}$
&  $\displaystyle{(-\eta)^{-3}}$         \\
\hline
%\begin{tabular}[c]{@{}c@{}}
%From bounce\\ to reentry
%\end{tabular}
\multicolumn{4}{|l|}{From bounce to reentry}\\
%\hline
 $\displaystyle{i\frac{4\sqrt{2}}{H_{ec}^2 a_{ec}^3}\Big((\frac{2}{H_{ec}a_{ec}})^{-3}-\frac{1}{2}(\frac{2}{H_{ec} a_{ec}}+\eta)^{-3}\Big)k^{-\frac{3}{2}}}$
& $\displaystyle{|1-\frac{1}{2}\frac{1}{(\frac{2}{H_{ec} a_{ec}}+\eta)^3}|}$
&$\displaystyle{-\frac{\sqrt{2}}{H_{ec}^2a_{ec}^3}(\frac{1}{3}-i(\frac{2}{H_{ec}a_{ec}}+\eta)^{-3}k^{-3})k^{\frac{3}{2}}}$
&     $\displaystyle{\eta^{-3}}$        \\
\hline
%\begin{tabular}[c]{@{}c@{}}
%After\\ horizon reentry
%\end{tabular}
\multicolumn{4}{|l|}{After horizon reentry}\\
%\hline
 $\displaystyle{-i\frac{1}{a(\eta)}\frac{3H_{ec}^3 a_{ec}^3}{4\sqrt{2}}cos(k\eta)k^{-\frac{7}{2}}}$
& $\displaystyle{\eta^{-2}}$
&$\displaystyle{\frac{1}{\sqrt{2k}}\frac{e^{ik\eta}}{a(\eta)}}$
& $\displaystyle{\eta^{-2}}$     \\
\hline
\end{tabular}
\caption{$\eta$-dependence of continuously and symmetrically matched tensor modes}
\label{Table_eta}
\vspace{0.25cm}
\end{table}
%%%% end of table of eta dependence .
%%% insert table on k-dependence here 
\begin{table}[]\footnotesize
\centering
\begin{tabular}{|c|c|c|c|}
\hline
k-dependence
& Symmetric-matching &Continuous-matching
&Inflation
 \\ \hline
\begin{tabular}[c]{@{}c@{}}Before\\ horizon exit\end{tabular}
& $k^{-\frac{1}{2}}$        & $k^{-\frac{1}{2}}$ & $k^{-\frac{1}{2}}$
  \\ \hline
\begin{tabular}[c]{@{}c@{}}From exit\\ to $\eta=0$\end{tabular}
& $k^{-\frac{3}{2}}$          & $k^{-\frac{3}{2}}$  & $k^{-\frac{3}{2}}$
\\ \hline
\begin{tabular}[c]{@{}c@{}}From $\eta=0$\\ to reentry\end{tabular}
& $k^{-\frac{3}{2}}$          & $k^{-\frac{3}{2}}$ & $k^{-\frac{3}{2}}$
  \\ \hline
\begin{tabular}[c]{@{}c@{}}After\\ horizon reentry\end{tabular}
 & $k^{-\frac{1}{2}}$         & $k^{-\frac{7}{2}}$ & $k^{-\frac{7}{2}}$
  \\ \hline
\end{tabular}
\caption{Scale dependence followed   from  two different  matching methods in the  simplified bounce, and  from the inflationary universe}
\label{Table_k_dependece}
\end{table}
\vspace{0.25cm}
%%% end of  table on k-dependence 

\subsection{The k-dependence in  the  power spectra}
We compared the k-dependence  followed   from  two different matching conditions  in the simplified 
CST bounce universe  with that of  the single-field  inflationary scenario~\cite{Alba:2015cms}. 
The results are presented in Table~\ref{Table_k_dependece}.

\section{From power spectra to observations}
\label{sec:power_spectrum}
Two complete power spectra -- one by imposing the continuity of wave-functions and their derivatives 
at the bounce point,  and the other one by insisting on the symmetry of the wave-functions about 
the bounce point -- are obtained in Section~\ref{sec:simplified_bounce} above. 
A few salient points of the power spectra are discussed  here. 
  The continuous  matching condition generates a  power spectrum identical to  the one from single-field 
inflation, as recently obtained by Alba and Maldacena~\cite{Alba:2015cms}, 
\begin{equation}
\label{k-dependence-simplified-bounce3}
\mathcal{P}_h=\frac{k^3|h|^2}{2\pi^2}\propto k^{-4},
\end{equation}
because $h$ is proportional to $k^{-\frac{7}{2}}$  (as  shown in Table~\ref{Table_eta}).
A  scale-invariant primordial power spectrum is generated  upon    
horizon exit; a  red-tilt  present-time spectrum  results  after   horizon reentry. 
One can compare~(Eq.~\ref{k-dependence-simplified-bounce3}) in the case of bounce 
with~(Eq.~\ref{power_inf_pgw}) in the case of inflation. 
On the other hand,  the symmetric matching condition  produces another  scale invariant primordial power spectrum, thanks also to the horizon exit mechanism.  
 The  present time power spectrum after horizon re-entry  is, however,  blue tilt, 
\begin{equation}
\label{power_spectrum_symmetric_present_time}
\mathcal{P}_h=\frac{k^3|h|^2}{2\pi^2}\propto k^2~.
\end{equation}
To compare the analytical results with cosmological observations,  we calculate the ratio of the tensor perturbations to the curvature perturbations and further compute the B-mode spectrum of Cosmic Microwave Background (CMB).

\subsection{The Tensor-to-Scalar Ratio}
To compare  the amplitude of tensor modes to the curvature perturbations 
one computes  the ``tensor-to-scalar ratio''  as defined by, % in~\cite{Quintin:2015rta},
\begin{equation}
\label{definition_r-t-s ratio}
r\equiv\frac{\mathcal{P}_h}{\mathcal{P}_\zeta},
\end{equation}
with the tensor modes power spectrum \(P_h\) given by (Eq.~\ref{k-dependence-simplified-bounce3}) 
 and the spectrum of the curvature perturbation defined as
\begin{equation}
\label{definition_curvature}
\mathcal{P}_\zeta\equiv\frac{k^3|\zeta_k|^2}{2\pi^2}.
\end{equation}
The  power spectrum of curvature perturbations in the  CSTB universe has been obtained 
in~\cite{Li:2013bha},    
\begin{equation}
\label{curvature_cstb}
\mathcal{P}_\zeta=(\frac{H}{\dot{T_c}})^2 \mathcal{P}_{\delta T}=\kappa^{-2} \mathcal{P}_{\delta T},
\end{equation}
where the spectrum of \(\delta T\) (\(P_{\delta T}\equiv\frac{k^3|\delta T_k|^2}{2\pi^2}\)) 
has been shown to be scale independent and stable against time  evolution.   
As Hubble parameter has  time dependence different k-mode exits the horizon at different times.  
Only  when this implicit time dependence is properly  taken in   account that one can establish the 
overall time dependence and scale dependence in the final power spectrum 
 of the scalar perturbations in the  metric~\cite{Li:2013bha, Li:2012vi, Ming:2017dtm}.
Let us recall that in the CST bounce universe  %the  time independence  on 
 perturbations of the matter field  \(\delta T\)  obey the equation at super-Hubble scales, 
\begin{equation}
\label{eq_deltat_cstb}
\ddot{\delta T_k}+\frac{k^2}{a^2}\delta T_k=0, 
\end{equation}
whose solution is 
\begin{equation}
\label{sl_deltat_cstb}
\delta T_k=C_{cur}\frac{1}{a(\eta)}\frac{e^{-ik\eta}}{\sqrt{2k}}(1-\frac{i}{k\eta})~.
\end{equation}
In  the discussion on the tensor modes above, we have set the BD vacuum as the initial condition
 in the far past.  The same 
%Likewise, to compare the amplitude of the tensor modes with scalar modes,   we  set the 
BD vacuum is set as the initial condition for  \(\delta T\), giving \(C_{cur}=\frac{1}{m_{pl}}\).    
Taking  \(k\eta\rightarrow+\infty\) in  (Eq.~\ref{sl_deltat_cstb})  we obtain curvature perturbations 
after the horizon reentry, 
\begin{equation}   
\label{curvature_cstb_afterreentry}   
\delta T_k=\frac{1}{m_{pl}}\frac{1}{a(\eta)}\frac{e^{-ik\eta}}{\sqrt{2k}}.
\end{equation}   
This leads  to  the tensor-to-scalar ratio,    
\begin{equation}   
r=\frac{\mathcal{P}_h}{\mathcal{P}_\zeta}=\frac{\kappa^2 \mathcal{P}_h}{\mathcal{P}_{\delta T}}~,   
\end{equation}    
where $\kappa$ is a dimensionful quantity defined by  
${\dot{T_{c}}}= \kappa H$~\cite{Li:2011nj, Li:2013bha}.
Reinstating the suppressed mass scales we have, 
\begin{equation}   
\label{ratio_kappa}   
r=\frac{\mathcal{P}_h}{\mathcal{P}_\zeta}=\kappa^2\, 
\frac{m_{s}^{2}}{m_{pl}^{2}} \, \frac{\mathcal{P}_h}{\mathcal{P}_{\delta T}}~,   
\end{equation}    
with  ${m_{s}}$ being  the string mass scale. 

\paragraph{The continuously matched power spectrum:}
The power spectrum  resulting  from the continuous matching condition can be read off in 
Table.~\ref{Table_eta}, 
\begin{equation}
\label{Power_spectrum_expression_con}
\mathcal{P}_{h(con)}=\frac{k^3}{2\pi^2}\frac{1}{a^2m_{pl}^2k^7}\frac{9H_{ec}^6a_{ec}^6}{32};
\end{equation}
whose  tensor-to-scalar ratio is thus, 
\begin{equation} 
\label{ratio_continuous}  
r_{con}= \frac{9}{16} \kappa^2 \frac{m_{s}^{2}}{m_{pl}^{2}} 
     \Big(\frac{H_{ec} a_{ec}}{k}\Big)^6.
\end{equation}  

\paragraph{The symmetrically matched power spectrum:}
In the symmetrically  matched  gravitational waves,  the power spectrum of tensor modes  after 
 horizon re-entry  is,  given  in Table.~\ref{Table_eta},  
\begin{equation}
\label{Power_spectrum_expression_sym}
\mathcal{P}_{h(sym)}=\frac{k^3}{2\pi^2 }\frac{1}{2a^2\, k}.
\end{equation}
And   the ratio of  ``tensor to scalar''  in the symmetrical case becomes,
\begin{equation}
\label{ratio_symmetrical}
r_{sym}=\kappa^2 \frac{m_{s}^{2}}{m_{pl}^{2}}~.
\end{equation}

Let us note here the quantity  \(\frac{H_{ec} a_{ec}}{k}\)  of  (Eq.~\ref{ratio_continuous})  
is  quite  big,  so is  the ratio of ``tensor to scalar''  (and the amplitude in Fig.~\ref{fig:subfig:a}) 
 in the  continuous case,  as compared to  analogous quantities  in  the symmetric case. 
As   \(\kappa\)   %% the definition of Kappa  
 is obtained from numerical solutions of the CSTB model and it is determined by 
the linear  relation between   the vacuum expectation  value of the tachyon field, $<T_{c}>$,
and Hubble parameter during the phase of tachyon matter dominated contraction, 
$<T_{c}>  = \kappa H $.  The numerical value of  \(\kappa\)  is pretty stable,  $1<  \kappa <10 $, 
over a wide range of parameter  values~\cite{Li:2013bha}.  
It  is then  unlikely to fiddle with the parametric values of the CSTB model to 
  suppress the high tensor to scalar ratio.   
Observations of tensor modes,  or the lack of it,  favours the small tensor to scalar ratio 
and hence the  symmetric gluing condition in a bounce universe.   
Once the tensor to scalar ratio is measured we can use it to put an upper bound on the ratio of  
string mass to the Planck mass.  
This small tensor to scalar ratio is still the Occam's  razor  for  alternatives to the inflationary paradigm. 

\subsection{Tensor modes of CSTB universe versus inflation}
\label{sec:comparison}  
\begin{figure}[htbp]
\centering
\includegraphics[width=0.8\textwidth]{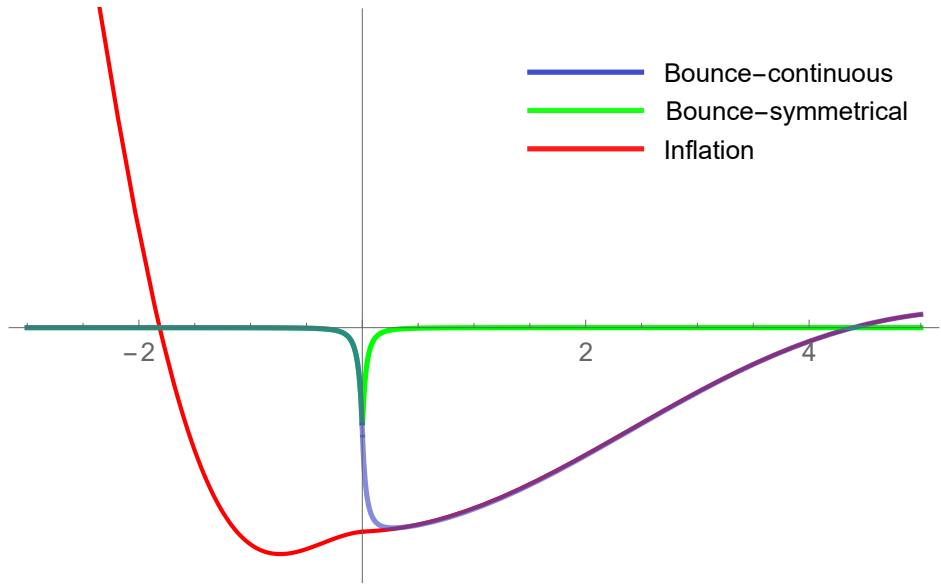}
\caption{The gravitational wave functions in the  three cases discussed in the paper. }
\label{fig_comparison}
\end{figure}

In this section we attempt to make a side by side comparison on  the similarities and the  differences  
of the gravitational  waves spectra  obtained from  the CST bounce universe model  
and  the single-field inflation  model~\cite{Alba:2015cms}.  %%%%
The complete GW wavefunctions %presented  in Fig.~\ref{fig:subfig:a},   
from   the bounce model  and  
%presented  in Fig.~\ref{fig:matching_inflation5} 
from  the single-field  inflation model   are plotted together in Fig.~\ref{fig_comparison}. 
The tensor modes of both matching conditions in  the bounce case  are  time-varying 
while the primordial tensor perturbations of the inflationary universe are 
time-invariant~\cite{Wands:1998yp}. The k-dependence in these three 
 cases are summarised  in Table~\ref{Table_k_dependece}. 
From Fig.\ref{fig_comparison}   one sees  that the continuously  matched 
bounce model shares a similar curve with the inflationary model 
after the horizon reentry, which explains the same k-dependence and 
the similar functional  forms of these two GW spectra.  
Detailed calculations relevant to the GW spectrum in the single-field inflation model~\cite{Alba:2015cms}  
is reproduced  in the Appendix~\ref{app:inflation} for easy reference. 

\subsection{The BB power spectrum of CMB}
The B-modes of CMB,  partly generated by the primordial gravitational waves, 
encodes the history of the universe earlier than the last scattering 
surface~\cite{Seljak:1996ti, Kamionkowski:1996zd, Seljak:1996gy}. 
In this paper we use the Code for Anisotropies in the Microwave Background (CAMB)~\cite{Lewis:1999bs} developed from CMBFAST~\cite{Seljak:1996is} 
to compute the BB power spectrum generated by the primordial GW in the  simplified 
CST bounce universe\footnote{The online tool of CAMB can be found at \url{https://lambda.gsfc.nasa.gov/toolbox/tb_camb_form.cfm.}}.

\begin{figure}
\centering
\subfigure[]
{
\label{fig:bbspectrum1} %% label for first subfigure
\includegraphics[width=0.48\linewidth]{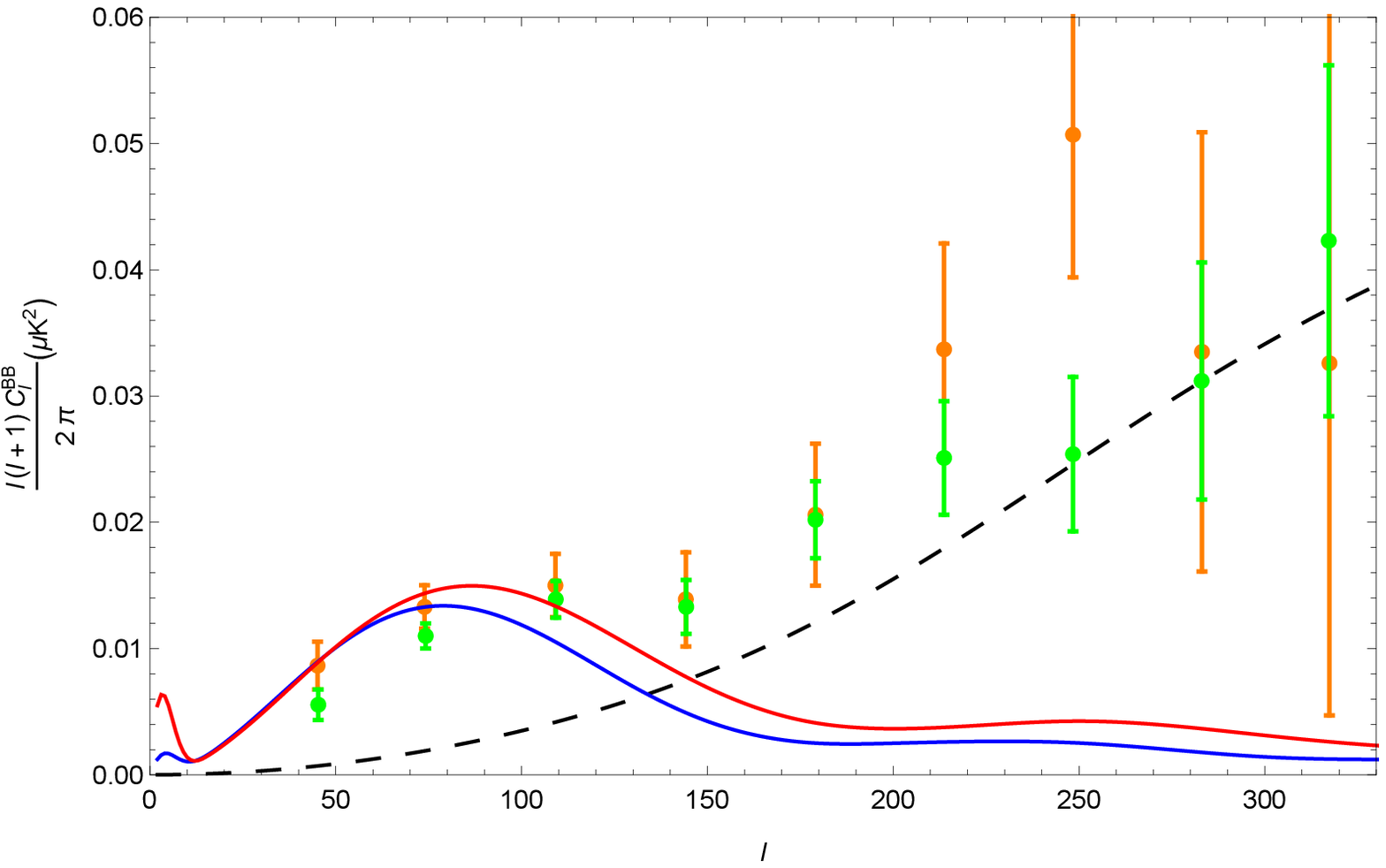}}
\subfigure[]
{
\label{fig:bbspectrum2} %% label for second subfigure
\includegraphics[width=0.48\linewidth]{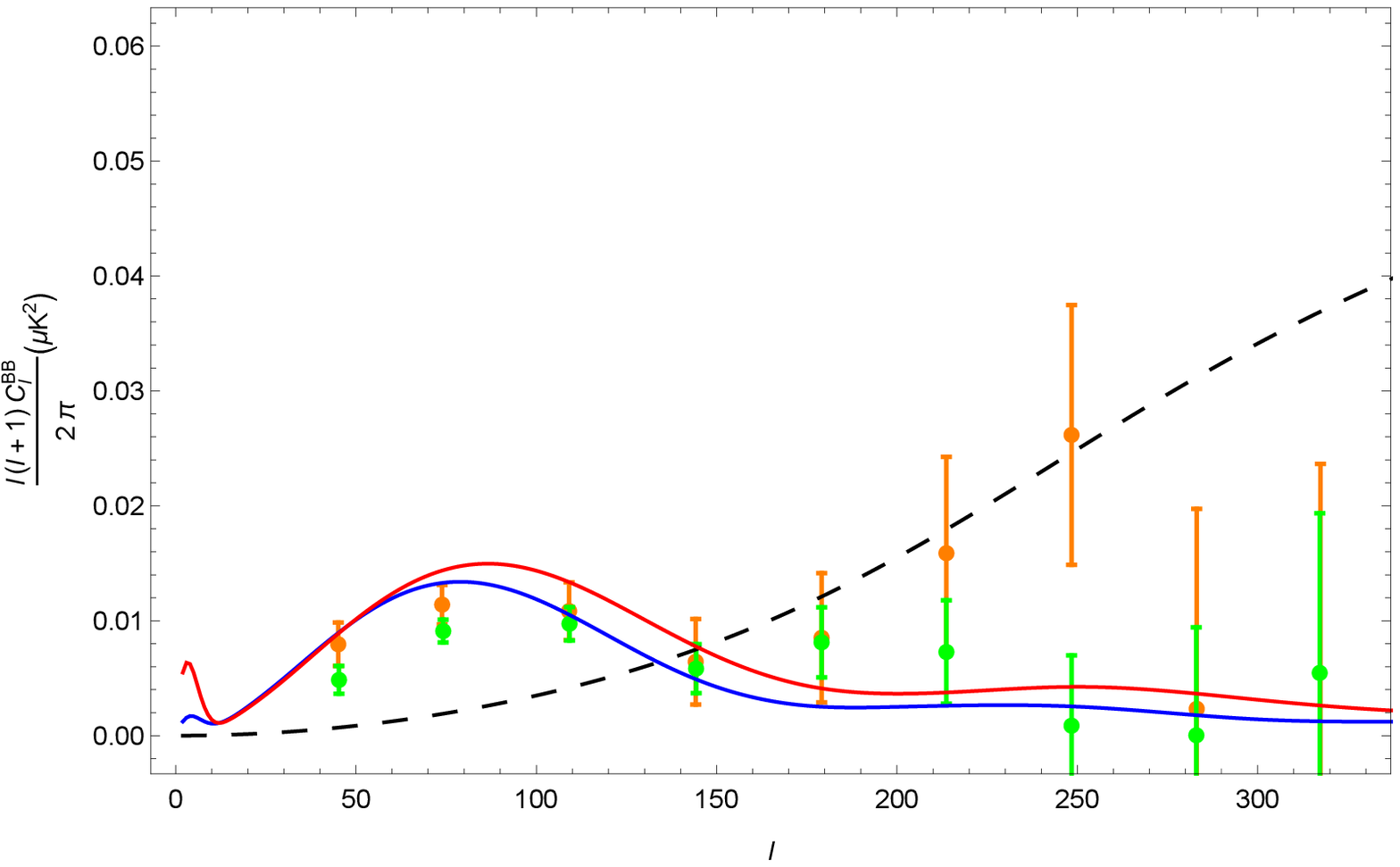}}
\caption{In both subfigures, the red curves,  and blue curves,  correspond to the BB 
power spectrum generated by the primordial gravitational waves of the  inflationary 
universe,  and the simplified CSTB universe,  respectively. 
The black dashed curve is the BB spectrum of the  lensed-\(\Lambda \)CDM. 
In Fig.~\ref{fig:bbspectrum1}, the orange dots, with error bars,  are BICEP2 data 
released in 2014~\cite{Ade:2014xna} while the green dots are data  released in 2015~\cite{Ade:2015fwj}. 
The dots inside Fig.~\ref{fig:bbspectrum2} are the data after 
subtracting the lense-\(\Lambda \)CDM  BB power spectrum part. 
This is done so to compare the ``pure'' gravitational waves contributions to the BB spectrum from both models.}
\label{fig:BBspectrum} %% label for entire figure 
\end{figure}

The BB power spectrum  is presented in~Fig.\ref{fig:BBspectrum},  which is plotted by setting the 
tensor mode  index  to zero, and the equation of state before horizon reentry to zero corresponding to 
the tachyon matter dominated contraction era.  
These are  done  in line with the  inflation model for a fair comparison. 
We have  also computed and plotted the B-modes in the single-field inflationary universe 
and the BB power spectrum of  the lensed-\(\Lambda \)CDM~\cite{Seljak:1996is}. 
We have set the tensor-to-scalar ratio to be 0.2 when computing the 
BB power spectrum generated by primordial GW.
From  Fig.~\ref{fig:BBspectrum}  one  finds  that the BB power spectrum in  the range 
(\(10<l<50\)) from the simplified CST bounce universe tracks closely in shape with  the  spectrum 
produced by the inflation model  and is consistently lower in amplitude.   

Considering the data of BB power spectrum of CMB and the lensed-\(\Lambda \)CDM  curve, 
the primordial power spectrum may be responsible for the B-modes of CMB 
at large angle when the ratio is around 0.2. 
However, with further detection and joint data analysis done by BICEP2 with Planck,  it was proposed that part of the B-mode signals was attributed to the Galactic dust and hence the ratio was further 
compressed~\cite{Ade:2018gkx, Adam:2014bub, Ade:2013aro, Hinshaw:2008kr, Ade:2015tva}. 
A   ``lensed-\(\Lambda\)CDM + tensor + dust''   framework  has since  been worked out, 
and the tensor-to-scalar ratio  has thus  been  suppressed to be lower than 0.07. 
We hope future detections  of the primordial gravitational waves, the like of LISA, Ali, and TianQin,   can  measure  with high precision  the primordial tensor perturbations.  
Only then one can hope to discern the fine details in the  gravitational waves spectra  
as predicted by the large repertoire  of early universe models.

\section{Conclusion}
In this paper we  have derived the  equations of motion  for the tensorial modes of the  metric 
perturbations  in a bounce universe whose dynamics is governed by the 
Coupled Scalar and Tachyon fields (CST) arising  in two stacks of colliding D-branes 
and anti-D-branes~\cite{Li:2011nj,Cheung:2016oab}.   
We obtain two sets of independent solutions in each of the following epochs: 
when the modes are  inside the horizon well before the bounce,  and well outside 
the horizon during the  matter dominated contraction phase characteristic of 
the bounce universe, and then finally when the gravitational waves re-enter the 
horizon while  the universe is in matter domination again. 
As gravitational waves are quite immune to the potentially complicated 
physics due to matter generation and subsequent developments  close to 
 the bounce  point due to its extremely weak coupling to matter fields,  
one can henceforth   simplify the bounce process. 
To  obtain the full wave functions of the gravitational waves throughout the 
whole cosmic  evolution and yet capture the interesting physics  pertinent to 
low-l modes of gravitational waves,  
%we  propose two reasonable matching  conditions: 
%a  continuous matching condition that insists on the continuity of the wave functions and their derivatives,  and a  symmetrical matching condition that respects the symmetry of the wave functions about the bounce point. 
we  propose two gluing conditions for the gravitational waves inside a bounce universe: 
 One insisting on the continuity of the wave functions and their derivatives at the 
 idealised bounce point; the other one respecting the symmetry of the bounce dynamics 
 about the bounce point. 
Both matching conditions predict scale invariance of the primordial spectrum  when the modes  are outside of the effective horizon,  and  oscillation modes  of the primordial tensor 
perturbations after the bounce. 
However, the continuous matching condition leads to  a much larger amplitude 
of the primordial tensor modes than  the symmetric matching condition.
The power spectrum of the primordial tensor perturbations of continuous matching 
is red tilt while that of symmetric matching becomes  blue in late time.
A detailed comparison is done with the primordial spectrum obtained from the single-field inflation 
model~\cite{Alba:2015cms}. 
We summary all features derived in this paper and their comparison with inflation  in  Table~\ref{Table_comparison}.
%% insert table comparison ... 
\begin{table}[]
\centering
\begin{tabular}{|c|c|c|c|}
\hline
Items   & Contiuous-matching & Symmetrical-matching & Inflation 
\\ \hline
\begin{tabular}[c]{@{}c@{}}Analytical form \\ 
of tensor modes\end{tabular}    & Table~\ref{Table_eta}      
&  Table~\ref{Table_eta} & Table~\ref{Table_inflation}                                                                      
\\ \hline
\begin{tabular}[c]{@{}c@{}}Primordial \\
Power Spectrum\end{tabular}  & 
\multicolumn{2}{|c|}{\begin{tabular}[c]{@{}c@{}}Scale-Invariant \\ 
Time-Varying\end{tabular}} & \begin{tabular}[c]{@{}c@{}}Scale-Invariant \\ 
Time-invariant\end{tabular}                                                                       
\\ \hline
\begin{tabular}[c]{@{}c@{}}Tensor-to-scalar \\
Ratio\end{tabular}  & $\kappa^2\frac{9}{16}\Big(\frac{H_{ec} a_{ec}}{k}\Big)^6$                  
& $\kappa^2$      & $2\epsilon$                                                                  
\\ \hline
\begin{tabular}[c]{@{}c@{}}B-modes \\ 
Curve\end{tabular}  & \multicolumn{3}{|c|}{Fig.\ref{fig:BBspectrum}}                                                                
\\ \hline
\end{tabular}
\label{Table_comparison}
\caption{Tensor modes comparison of simplified CSTB universe and inflationary cosmos}
\vspace{0.25cm}
\end{table}
%% end of  table comparison ... 
 
%%
Higher precision  measurements of BB power spectrum in the Cosmic Microwaves Background 
can place better constraints on the inflationary paradigm and its alternatives. 
Direct detections of primordial gravitational waves from LISA, ALI, or TianQin and 
the future array of low-frequency GW detectors can eventually tell whether our universe 
went through a bounce or an inflation.  

Meanwhile particle physics experiments aiming at direct detections of dark matter and 
determination of their coupling with Standard Model particles are in full swing.  
As bounce universe allows an out of thermal equilibrium route of dark matter creation in the contraction phase, dark matter mass and their coupling are constrained by the present-day relic 
abundance~\cite{%
Yeuk-KwanEdnaCheung:2016zra, Vergados:2016niz, Cheung:2014pea}. 
One may be in for surprise earlier than expected, as it provides a way to distinguish 
inflation~\cite{Guth:1980zm} from a bounce universe.

%%%%%%%%%%%%%%%%%%%%%%%%%%%%%%%%%%%%%%
%%%%  APPENDIX
%\input{appendix.tex}

%%%%%%%%%%%%%%%%%%%%%%%%%
%%%%  APPENDIX
%%%%%%%%%%%%%%%%%%%%%%%%%%%%%
\appendix
\section{The primordial gravitational waves in the Standard Inflationary  Cosmology}
\label{app:inflation}
The primordial power spectrum of the gravitational waves arisen in the inflation model is obtained recently  by Alba and Maldacena~\cite{Alba:2015cms}.
We review their  calculations in detail here  for pedagogical purpose and for quick comparison with our results presented in the main body of our paper.
%of tensor modes of inflationary universe applying the similar method discussed in the section of simplified bounce universe.
\subsection{The single field slow-rolling inflation and the evolution of the scale factor}
\label{Asec:inflation-scale-factor}
Based on Einstein's theory of relativity,  one can construct a model to realise inflation with a simple scalar field Lagrangian,
$\mathscr{L}(\phi)=-\partial^\mu\phi\partial_\mu\phi-V(\phi)$.
The  energy density and the pressure in the stress-energy tensor are
\begin{equation}
\label{inflation_density}
\rho=\frac{1}{2}\dot\phi^2+V(\phi)
\end{equation}
and
\begin{equation}
\label{inflation_pressure}
p=\frac{1}{2}\dot\phi^2-V(\phi)~,
\end{equation}
will be driving an exponential expansion of the underlying universe provided that 
 the EoS (Equation of State) 
\begin{equation}
\label{inflation_eos_general}
\omega=\frac{p}{\rho}=\frac{\dot{\phi}^2-2V(\phi)}{\dot{\phi}^2+2V(\phi)}
\end{equation}
can stay $\omega=-1$  for long enough time. 
The easiest is to impose the slow-roll conditions~\cite{Linde:1981mu}:
$$
 \epsilon_{V} \equiv \frac{1}{16\pi G_{N}} (\frac{ V^{\prime}}{V})^{2} \ll 1
$$
and 
$$
\eta_{V} \equiv =  \frac{1}{ 8 \pi G_{N}} \frac{V^{\prime\prime}}{V}  \ll 1~.
$$
In other words, the kinetic energy of the inflaton is negligible in comparison to 
its potential energy and its potential hill being ultra-flat. 

With %the equation of state of the universe \(\omega\) being constant and
an almost  constant  Hubble parameter  during an inflation,  one can obtain an analytical expression of the scale factor  by solving the Friedmann equations:
\begin{equation}
\label{Aeq:a_inflation_t}
a(t)=\left\{\begin{array}{lcl}
\begin{aligned}
&a_{\Lambda}e^{H_\Lambda t},\  &t<0\\
&a_{\Lambda}(1+\frac{H_\Lambda t}{n})^{n},\ &t\ge0
\end{aligned}
\end{array}\right.
\end{equation}
where the zero time \(t=0\) is set at  the time when inflation ends and $a_{\Lambda}$ 
denotes the scale factor of the universe attained at the end of inflation.  
The minus and the positive branches of the scale factor evolution  correspond to the 
inflationary stage and the post-inflation expansion respectively. 
The index $n$ in the power law expansion after inflation is determined  by  
the EoS of the cosmos, $n=\frac{2}{3(1+\omega)}$.
For the study of gravitational waves in an inflationary universe we recall a few key 
details of the single field inflation model below. 

The scale factor of the inflationary universe~(Eq.~\ref{Aeq:a_inflation_t}) and its 
time derivative remains continuous during the whole cosmological evolution.  
The physics at the end of the inflation is however complicated due to matter production: ``reheating'' is    a process  in which matter is purported to be produced 
from the decay of the inflaton. For a recent discussion on the interesting physics 
that can happen during  reheating and its implications on reheating temperature, 
the readers  are referred to~\cite{Drewes:2015coa}.
In fact one can  turn this around and ask how %%%%%
 reheating and parametric resonance can have imprints on 
CMB;  and whether CMB can tell us anything about the highest temperature the universe 
has ever reached in the earliest epoch of evolution.  
For the present discussion, however,    ``reheating''  and the interactions of matter occurred thereafter are completely  ignored, as they play no role in the cosmic evolution 
around the time when tensor modes exit the Hubble horizon. 

When the tensor modes cross the horizon during the inflation and in the 
post-inflation expansion,  the analytical form of the scale factor is  given 
by~(Eq.~\ref{Aeq:a_inflation_t}).
In conformal time $\eta=\int\frac{dt}{a(t)}$,  the scale factor takes the form: 
\begin{equation}
\label{Aeq:a_inflation_eta}
a(\eta)=\left\{\begin{array}{lcl}
\begin{aligned}
&\frac{a_{0} }{1-H_\Lambda a_0 \eta},\  &\eta<0\\
&a_{0}\Big(1+\frac{H_\Lambda a_0 \eta}{\nu}\Big)^{\nu},\ &\eta\ge0
\end{aligned}
\end{array}\right.
\end{equation}
with $\nu=\frac{2}{1+3\omega}$.

\subsection{The equation of linearised  tensorial  perturbations of the background metric in the single-field  inflation model}
The isotropic homogeneous background universe is  described by the FLRW metric
\begin{equation}
\label{Aeq:FRLW} 
g_{\mu\nu}^{(0)} = diag\{-1,a(t)^2,a(t)^2,a(t)^2\}~,
\end{equation}
with the time variable $t$ being the physical time. 
The tensor perturbations can be expressed as 
$g_{\mu\nu} = g^{(0)}_{\mu\nu} + \delta g_{\mu\nu}$.
The metric tensor perturbations are traceless and divergence-less,
\begin{equation}
\label{Aeq:traceless-divergenceless} 
\delta g_{\mu\nu} \equiv h_{\mu\nu}: ~~  h^i_i=h^{;i}_{ij}=0~.
\end{equation}
There are thus   two  independent spin-2   tensor modes of metric perturbations,  
\begin{equation}  
 \label{Aeq:perturbed-metric1}
\delta g_{\mu\nu}=
a(t)^2  \left(\begin{array}{lcl}
         \begin{aligned}
          & 0 \quad & 0 \qquad 		   &  0  \qquad   &  0     \\
          & 0 \quad & h_{\times}\qquad &  h_+ \qquad  &  0   \\
          & 0 \quad & h_{+} \qquad	   & -h_{\times}  &  0  \\
          & 0 \quad & 0 \qquad         &   0  \qquad  &  0
         \end{aligned}
\end{array}\right)~,
\end{equation}
describing the propagation of  gravitational waves  at the speed of light 
in the $ \hat{k}=\hat{z}$ direction.

The equations of tensor perturbations can be obtained by expanding the action,
\begin{equation}
\label{Aeq:Infl-action}
S=\int\sqrt{-g}\big(\mathscr{L}_{inf}(\phi+\delta\phi)
-\frac{R(g^{(0)}_{\mu\nu}, \delta g_{\mu\nu})}{16\pi G}\big)d^4x,
\end{equation}
The determinant of the metric,  $\sqrt{-g}$, and the Ricci scalar, $R$,  
can be expanded  up to the quadratic order in $h_{ij}$ as follows, 
\begin{equation}
\label{Aeq:SQRT-g} 
\sqrt{-g}=a^3(1-\frac{1}{4}h_{ij}h_{ij}),
\end{equation}
and
\begin{equation}
\label{Aeq:Ricci}
R=6(H^2+\frac{\ddot{a}}{a})-\frac{3}{4}\dot{h_{ij}}^2-h_{ij}\ddot
h_{ij}-\frac{1}{4a^2}h^2_{ij,k}-4Hh_{ij}\dot{h_{ij}},
\end{equation}
where the dot  denotes derivative  with respect to the physical time. 
In the derivation of~(Eq.~\ref{Aeq:SQRT-g}) and~(Eq.~\ref{Aeq:Ricci}), 
the condition~(Eq.~\ref{Aeq:traceless-divergenceless}) has
been used.  Roman indices ${i,j,k}$ run over {1, 2, 3}.

Expanding the action~(Eq.~\ref{Aeq:Infl-action})  to the quadratic order and  by 
 variational method we obtain the equation of tensor perturbations of the metric,
\begin{equation}
\label{Aeq:infl-tensor}
\ddot{h}_{ij}+ 3 H \dot{h}_{ij}-\Delta h_{ij}=0~.
\end{equation}
%with the help of Einstein equation of the homogenous and isotropic background field, 
Since we are interested in the full spectrum of gravitational waves, we need 
the equation  of motion for  $h$ in Fourier space,
\begin{equation}
\label{Aeq:infl-tensor1}
{h}_{ij}^{\prime\prime}+2\frac{a^{\prime}}{a}{h}_{ij}^{\prime}+k^2 h_{ij}=0~,
\end{equation}
where we have changed the time variable to  conformal time, $\eta$,  and  prime 
 denotes a derivative  with respect to $\eta$. Solving~(Eq.~\ref{Aeq:infl-tensor1}), 
we obtain  the evolution of   gravitational waves spectrum.

\subsection{Solving the equation of tensor modes in a single-field  inflation model}
When  the scale factor expands with  a  power law $a\propto \eta^\nu$, 
with the  substitution $u=ah$,  (Eq.~\ref{Aeq:infl-tensor1})~can 
always be transformed to the form,
\begin{equation}
\label{Aeq:eq_u_nu_geneal}
u^{\prime\prime}+\Big(k^2-\frac{\nu(\nu-1)}{\eta^2}\Big)u=0~.
\end{equation}
The solution of $u(\eta)$ is  a linear combination of 
$\eta^{\frac{1}{2}}J_{\nu-\frac{1}{2}}(k\eta)$ 
and $\eta^{\frac{1}{2}}J_{\frac{1}{2}-\nu}(k\eta)$~\footnote{%
We have ignored a special case of $\nu=\frac{3}{2}$, where the solution becomes a 
combination of Hankel functions. The result is the same as the $\nu\neq\frac{3}{2}$ case.}.
Applying the scale factor of inflationary universe~(Eq.~\ref{Aeq:a_inflation_eta})
into~(Eq.~\ref{Aeq:eq_u_nu_geneal}), the solution to the tensor modes can be expressed  as,
\begin{equation}
\label{Aeq:h_solution_simple_inflation}
\footnotesize
\left\{\begin{array}{lcl}
\begin{aligned}
&h=\frac{C_{inf}}{a(\eta)}\Big(\frac{1}{H_\Lambda a_0}-\eta\Big)^{\frac{1}{2}}J_{\frac{3}{2}}(\frac{k}{H_\Lambda a_0}-k\eta)
+\frac{D_{inf}}{a(\eta)}\Big(\frac{1}{H_\Lambda a_0}-\eta\Big)^{\frac{1}{2}}J_{-\frac{3}{2}}(\frac{k}{H_\Lambda a_0}-k\eta),&\eta<0\\
&h=\frac{C_{exp}}{a(\eta)}\Big(\frac{\nu}{H_\Lambda a_0}+\eta\Big)^{\frac{1}{2}}J_{\nu-\frac{1}{2}}(\frac{k\nu}{H_\Lambda a_0}+k\eta)
+\frac{D_{exp}}{a(\eta)}\Big(\frac{\nu}{H_\Lambda a_0}+\eta\Big)^{\frac{1}{2}}J_{\frac{1}{2}-\nu}(\frac{k\nu}{H_\Lambda a_0}+k\eta),&\eta\ge0\ .\\
\end{aligned}
\end{array}
\right.
\end{equation}
The two branches of the solutions in~(Eq.\ref{Aeq:h_solution_simple_inflation})  
correspond to   inflation \((\eta<0)\) and the post-inflationary expansion 
\((\eta>0)\). They are connected at the end of  inflation \((\eta=0)\).
Typically, the coefficients ($C_{inf},D_{inf}$) of the wave functions during inflation 
are determined by initial conditions;  and the coefficients ($C_{exp},D_{exp}$) 
of the wave functions during the post-inflation expansion are then related to ($C_{inf},D_{inf}$) at the end of  inflation.

\paragraph{The horizon crossing conditions of the tensor modes:  }
The horizon crossing behaviour of the tensor modes during and after the inflation 
is similar to the analogous horizon crossing process in  the  bounce universe, 
the latter  has been discussed  in Sec.~\ref{sec:horizon_cross} and we refer to 
D.~Wands's original paper~\cite{Wands:1998yp} for more   details. 
Thus applying   the Equation of State  at  horizon crossing~(Eq.~\ref{horizon_crossing_condition3}), 
one obtains  the condition for a given $k$-mode of gravitational  wave to  exit 
  the horizon,
\begin{equation}
\label{Aeq:inflation_horizon_exit}
\frac{k}{aH}=k\eta=-1
\end{equation}
and reenter the horizon,
\begin{equation}
\label{Aeq:inflation_horizon_reentry}
k\eta=\nu.
\end{equation}
The horizon crossing conditions hence divide the gravitational wave function 
into three stages. The inequality \(k|\eta|<|\nu|\) (\(\eta\rightarrow0\)) 
corresponds to physical modes being well outside the horizon, 
when at large time limit (\(\eta\rightarrow \infty\))  all  k-modes being
 well inside the horizon. 
In summary,
\begin{equation}
\label{Aeq:k_eta_inflation}
\left\{\begin{array}{lcl}
\begin{aligned}
&k\eta\rightarrow -\infty,\ & \mathrm{in\,  the\,  far\, past; }\\
&k\eta\rightarrow 0,      \ &\mathrm{outside\, the\, horizon;} \\
&k\eta\rightarrow +\infty,\ & \mathrm{in \, the \, far\, future.}\\
\end{aligned}
\end{array}\right.
\end{equation}
which is the same as in the bounce case~(Eq.~\ref{eq:k_eta_simplified_bounce}).

\paragraph{The short wavelength and long wavelength limits:  }
Applying the limits~(Eq.~\ref{Aeq:k_eta_inflation}) to the general 
solutions~(Eq.~\ref{Aeq:h_solution_simple_inflation}), one  obtains 
the expression for each k-mode in each of the following situation:
\begin{itemize}
\item
During inflation~($\eta<0$), the k-mode is well inside the
horizon~($\eta\rightarrow-\infty$):
\begin{equation}
\label{h_inf_contracting_inside}
h=-\frac{C_{inf}}{a(\eta)}\sqrt{\frac{2}{\pi k}}cos(\frac{k}{H_\Lambda a_0}-k\eta)-\frac{D_{inf}}{a(\eta)}\sqrt{\frac{2}{\pi k}}sin(\frac{k}{H_\Lambda a_0}-k\eta).
\end{equation}
\item
During inflation~($\eta<0$), the k-mode is far outside the
horizon~($\eta\rightarrow 0^{-}$):
\begin{equation}
\label{h_inf_contracting_outside}
h=C_{inf}H_\Lambda\frac{1}{3}\sqrt{\frac{2}{\pi}}k^{\frac{3}{2}}(\frac{1}{H_\Lambda a_0}-\eta)^3
-D_{inf}H_\Lambda \sqrt{\frac{2}{\pi}}k^{-\frac{3}{2}}
\end{equation}
\item
During the post-inflation expansion~($\eta> 0$), the k-mode is far outside the
horizon~($\eta\rightarrow 0^{+}$):
\begin{equation}
\label{h_inf_expansion_outside}
h=\frac{C_{exp}\nu^\nu}{H_\Lambda^\nu a_0^{\nu+1}}\frac{2^{\frac{1}{2}-\nu}}{\Gamma(\frac{1}{2}+\nu)}k^{\nu-\frac{1}{2}}
+\frac{D_{exp}\nu^\nu}{H_\Lambda^\nu a_0^{\nu+1}}\frac{2^{\nu-\frac{1}{2}}}{\Gamma(\frac{3}{2}-\nu)}k^{\frac{1}{2}-\nu}(\frac{\nu}{H_\Lambda a_0}+\eta)^{1-2\nu}
\end{equation}
\item
During the post-inflation expansion~($\eta> 0$), the k-mode is well inside  the
horizon~($\eta\rightarrow +\infty $):
\begin{equation}
\label{h_inf_expansion_inside}
h=\frac{1}{a(\eta)}\sqrt{\frac{2}{\pi k}}\Big(C_{exp}cos(k\eta-\frac{\nu\pi}{2})-D_{exp}sin(k\eta+\frac{\nu\pi}{2})\Big),
\end{equation}
\end{itemize}
with \(a(\eta)\) being the scale factor of the inflationary cosmos given 
by~(Eq.~\ref{Aeq:a_inflation_eta}).

\paragraph{Bunch-Davies vacuum:}
The universe is expected  to  start from a time-independent ground state in the 
far past.  A natural choice of the initial condition of the universe is the 
Bunch-Davies vacuum,
%%%
\begin{equation}
\label{BD_vacuum}
h=\frac{1}{m_{pl}}\frac{1}{\sqrt{2k}}\frac{e^{-ik\eta}}{a(\eta)},
\end{equation}
%%%
where $m^{2}_{pl}\equiv \frac{1}{8\pi G_{N}}\equiv\,1$ is the reduced Plank mass. 
The $\frac{1}{\sqrt{2k}}$ term arises due to the normalisation of the wave function,  
$|u|^2(2k)=1$, with  
$$
u_k(\eta)=a(\eta)h_k m_{pl}e^{-ik\eta}
$$
being  regarded as a quantized field.

Combining the solution in far past~(Eq.~\ref{h_inf_contracting_inside}) with the 
initial condition~(Eq.~\ref{BD_vacuum}), the coefficients during inflation are therefore 
determined,
\begin{equation}
\label{infl_initial_coeffiecients}
\left\{\begin{array}{lcl}
\begin{aligned}
&C_{inf}=-\frac{\sqrt{\pi}}{2m_{pl}}~,\\
&D_{inf}=i\frac{\sqrt{\pi}}{2m_{pl}}~.
\end{aligned}
\end{array}\right.
\end{equation}
Once this is done the  coefficients of the tensor perturbations during the 
post-inflationary  epoch  are also determined  using the matching 
conditions on the super-Hubble scale.

\subsection{Matching conditions}
\begin{figure}
\centering
\subfigure[]
{
\label{fig:inf:a} %% label for first subfigure
\includegraphics[width=0.48\linewidth]{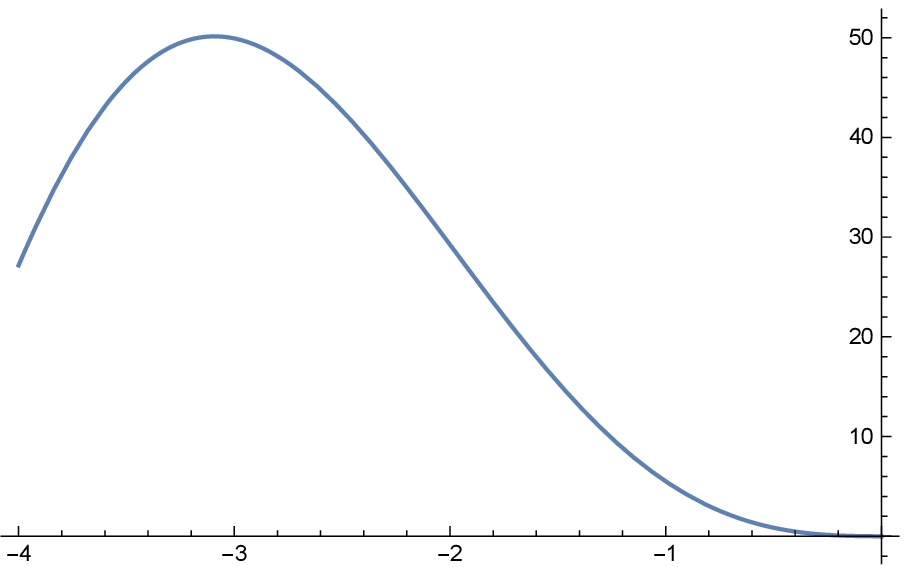}}
\subfigure[]
{
\label{fig:inf:b} %% label for second subfigure
\includegraphics[width=0.48\linewidth]{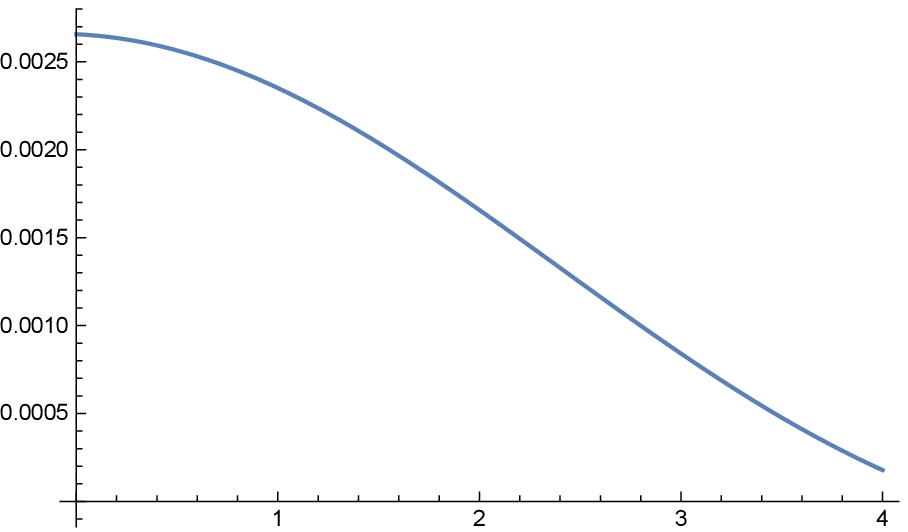}}
\subfigure[]
{
\label{fig:inf:c} %% label for first subfigure
\includegraphics[width=0.48\linewidth]{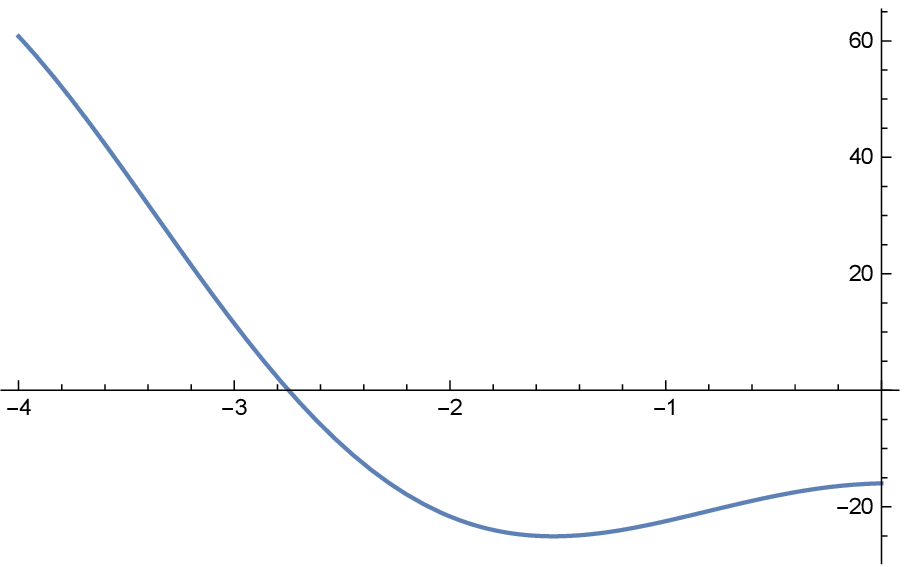}}
\subfigure[]
{
\label{fig:inf:d} %% label for second subfigure
\includegraphics[width=0.48\linewidth]{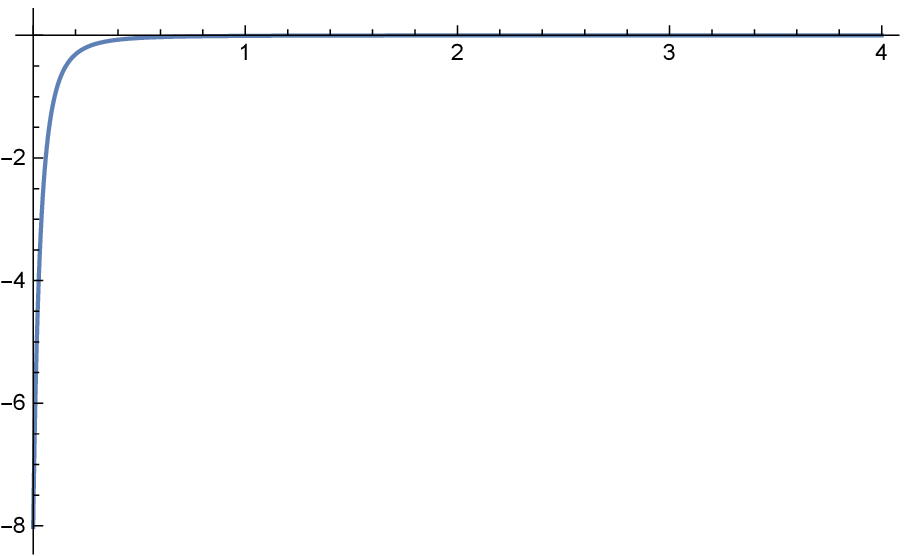}}
\caption{The two  branches to be matched at the end of the inflation. 
The subfigures Fig.~\ref{fig:inf:a} and Fig.~\ref{fig:inf:c} on the left side 
are the perturbations during the inflation while the two on the right side are the 
branches during the post-inflationary epoch of expansion. 
The task of matching is to connect a linear combination of the two functions 
on the right to a particular combination of the functions on the left
given by  (\(C_{inf},D_{inf}\)). 
The plots are generated with  $H_\Lambda=20$, $a_0=1$, $k=1$ and $\nu=2$. }
\label{fig:inf} %% label for entire figure \end{figure}
\end{figure}

\begin{figure}[htbp]
\centering
\includegraphics[width=0.8\textwidth]{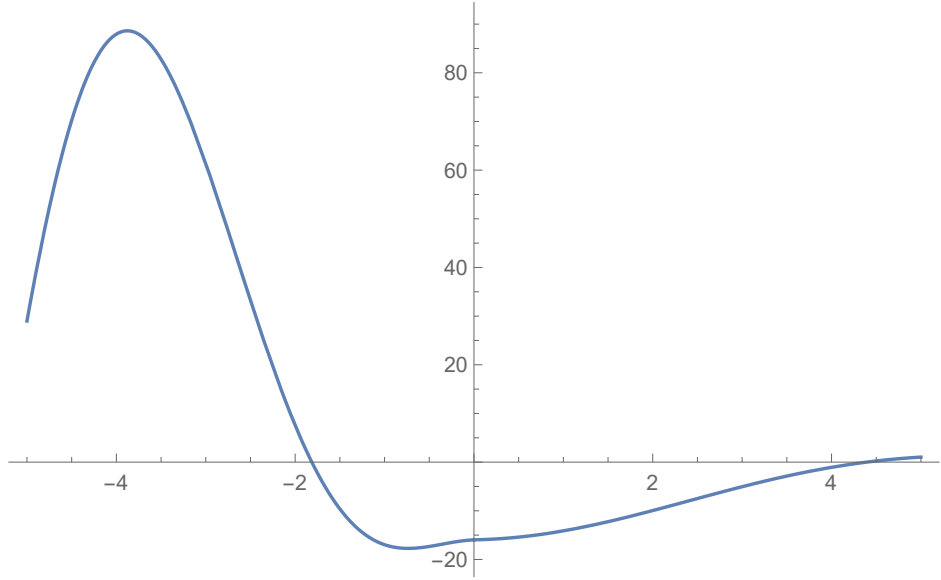}
\caption{Matched tensor modes near the end of the inflation. 
In the simulation,  the super-Hubble scale part is (\(-1\sim2\)) along the 
horizontal axis. It is obviously that during this period the k-mode is nearly 
frozen  (i.e. time-independent) as the function does not vary significantly 
 in the  epoch (\(-1<\eta<2\)).}
\label{fig:matching_inflation5}
\end{figure}

It is a natural  to assume  that the gravitational waves and their derivatives  
be continuous during the cosmological evolution.  The continuity of 
\(h(\eta)\) and \(h'(\eta)\) thus  relate   the post-inflation 
coefficients (\(C_{exp},D_{exp}\)) to (\(C_{inf},D_{inf}\)) at 
\(\eta\rightarrow0\) by,
\begin{equation}
\label{matching_conditions_inflation2}
\left\{\begin{array}{lcl}
\begin{aligned}
&C_{exp}=C_{inf}\frac{2^\nu\Gamma(\frac{1}{2}+\nu)}{\sqrt{\pi}}(\frac{1}{3\nu^\nu}-\frac{\nu^{1-\nu}}{2\nu-1})(\frac{k}{H_\Lambda a_0})^{2-\nu}
-D_{inf}\frac{\Gamma(\frac{1}{2}+\nu)}{\sqrt{\pi}}(\frac{2}{\nu})^\nu(\frac{k}{H_\Lambda a_0})^{-1-\nu},\\
&D_{exp}=C_{inf}\frac{2^{1-\nu}\nu^\nu\Gamma(\frac{3}{2}-\nu)}{(2\nu-1)\sqrt{\pi}}(\frac{k}{H_\Lambda a_0})^{\nu+1}.
\end{aligned}
\end{array}\right.
\end{equation}
Regardless of the constant terms, the matching condition acts to  transfer  
$\frac{k}{H_\lambda a_0}$ between coefficients. 
Since the tensor modes are stretched outside of the horizon near the end of 
the inflation, the long wavelength limit $\frac{k}{H_\Lambda a_0}\rightarrow0$ 
makes it possible to compare the contributions of ($C_{inf},D_{inf}$) to 
 ($C_{exp},D_{exp}$).  Obviously, the $C_{inf}$ term is 
$(\frac{k}{H_\lambda a_0})^3$ smaller than the $D_{inf}$ term in $C_{exp}$. 
The matched wave functions  are presented  numerically,  with $H_\Lambda=20$, 
$a_0=1$, $k=1$ and $\nu=2$,  in Fig.~\ref{fig:inf} 
and Fig.~\ref{fig:matching_inflation5}.

%%%
%%%%

\subsection{The power spectrum}
Combining  the coefficients during the inflation and post-inflationary expansion 
and the solutions of every cosmic  stage, the analytical form of each  epoch is 
summarized in Table.~\ref{Table_inflation}.
\begin{table}[]
\centering
\caption{Analytical form of the tensor perturbations in the single field 
inflationary universe.}
\label{Table_inflation}
\begin{tabular}{|c|c|c|c|}
\hline
Inflation   & Analytical solution & $\eta$-dependence & k-dependence 
\\ \hline
Far past    & $\frac{1}{m_{pl}}\frac{1}{\sqrt{2k}}\frac{e^{-ik\eta}}{a(\eta)}$                                                             & $1-H_\Lambda a_0 \eta$     &    $k^{-\frac{1}{2}}$     
\\ \hline
\begin{tabular}[c]{@{}c@{}}Super-Hubble\\ \(\eta\rightarrow0^-\)\end{tabular}    
& $h=-\frac{H_\Lambda k^{-\frac{3}{2}}}{m_{pl}\sqrt{2}}\Big(i+\frac{1}{3}(\frac{k}{H_\Lambda a_0}-k \eta)\Big)$                                                             
& $(-\eta)^{0}$      &  $k^{-\frac{3}{2}}$     
\\ \hline
\begin{tabular}[c]{@{}c@{}}Super-Hubble\\ \(\eta\rightarrow0^+\)\end{tabular} 
& $h=-\frac{H_\Lambda k^{-\frac{3}{2}}}{m_{pl}\sqrt{2}}
\Big(i+\frac{\nu^{2\nu}}{2\nu-1}(\frac{k}{H_\Lambda a_0})^{2\nu+2}
(\frac{k\nu}{H_\Lambda a_0}+k\eta)^{1-2\nu}\Big)$
& $\eta^{0}$     &     $k^{-\frac{3}{2}}$    
\\ \hline
Far future  & $h=-i\frac{2^\nu}{\sqrt{2\pi}}\frac{\Gamma(\frac{1}{2}+\nu)}{m_{pl}}\frac{H_\Lambda}{k^{\frac{3}{2}+\nu}}\frac{cos(k\eta-\frac{\nu\pi}{2})}{(\frac{\nu}{H_\Lambda a_0}+\eta)^\nu}$                                                             
& $(\frac{\nu}{H_\Lambda a_0}+\eta)^{-\nu}$  
&  $k^{-(\frac{3}{2}+\nu)}$    
\\ \hline
\end{tabular}
\end{table}
The primordial power spectrum becomes
\begin{equation}
\label{power_inf_pgw}
\mathcal{P}_h=\frac{k^3|h|^2}{2\pi^2}\propto k^0,
\end{equation}
while the power spectrum after the horizon reentry is
\begin{equation}
\label{power_inf_pgw}
\mathcal{P}_h=\frac{k^3|h|^2}{2\pi^2}\propto k^{-4},
\end{equation}
of which the EoS of the expansion phase during the horizon reentry is set  
at  \(\nu=2\) for a matter-dominated epoch.

\paragraph{The tensor-to-scalar ratio}
The readers are referred to~\cite{Quintin:2015rta} for a quick discussion on the calculations of scalar modes and the tensor-to-scalar ratio.
Employing  the Sasaki-Mukhanov variable \(v\), 
\begin{equation}
v\equiv z\zeta
\end{equation}
on super-Hubble scales, while the curvature perturbation \(\zeta\) observes
\begin{equation}
\label{eq_scalar}
\zeta_k^{\prime\prime} +2\frac{z'}{z}\zeta_k^{\prime} +k^2\zeta_k=0,
\end{equation}
with \(z=a\frac{\dot{\phi}}{H}m_{pl}\) in single field inflation model and being 
expressed by the slow-roll parameter \(z=am_{pl}\sqrt{2\epsilon}\). 
We assume that the initial condition of the curvature perturbation \(\zeta\) become 
\(\zeta=\frac{1}{z}\frac{e^{-ik\eta}}{\sqrt{2k}}\) and thus the ratio becomes
\begin{equation}
r=\frac{|h_k|^2}{|\zeta_k|^2}=(\frac{z}{a m_{pl}})^2=(\frac{\dot{\phi}}{H})^2=2\epsilon~.
\end{equation}
predicting almost negligible gravitational waves productions from single-field 
inflation model.

\acknowledgments
We would like to thank Changhong Li for many useful discussion at various stages 
of the project.
We would also thank  Zhong-Kai Guo (ITP, Beijing) and Yiqiu Ma (CalTech) for discussion.

This research project has been supported in parts by the NSF China
under Contract No.~11775110, No.~11690034. % and No.~11405084.
We also acknowledge the European Union's Horizon2020 research and innovation
programme (RISE) under the Marie Sk\'lodowska-Curie  grant agreement
No.~644121, and  the Priority Academic Program Development for
Jiangsu Higher Education Institutions (PAPD).

\paragraph{[Note added in proof:]}  After the completion of the draft,  we were informed by Changhong Li that the scalar to tensor ratio of the CST bounce universe has been obtained by *** in his master 
thesis supervised by Changhong Li~\cite{Feng:2018abc}.   
In this thesis a symmetric gluing condition was implicitly assumed.

%%%%%%%  use bib file %%%%%

\addcontentsline{toc}{section}{References}
\bibliographystyle{JHEP}
%\bibliographystyle{apj}

%%%%%%%  using \bibitem %%%%%%%%%%%%%%%%%%%%
%\bibliography{GW_ZC}
\providecommand{\href}[2]{#2}\begingroup\raggedright\endgroup

\end{document}